\documentclass[useAMS,usenatbib]{mn2e}

\newcommand{\be} {\begin{equation}}

\newcommand{\ee} {\end{equation}}

\newcommand{\INT}{{\it INTEGRAL}\,}

\newcommand{\CXO}{{\em Chandra}}
\newcommand{\R}{{\em ROSAT}}

\newcommand{\xmm}{{\em XMM--Newton}}
\newcommand{\XMM}{{\em XMM--Newton}}

\newcommand{\RXTE}{{\em RXTE}}
\newcommand{\swift}{{\em Swift}}
\newcommand{\su}{{\em Suzaku}}
\newcommand{\bc}{\begin{center}}
\newcommand{\ec}{\end{center}}
\def\ltsima{$\; \buildrel < \over \sim \;$}
\def\lsim{\lower.5ex\hbox{\ltsima}}
\def\loe{\lower.5ex\hbox{\ltsima}}
\def\gtsima{$\; \buildrel > \over \sim \;$}
\def\gsim{\lower.5ex\hbox{\gtsima}}
\def\goe{\lower.5ex\hbox{\gtsima}}
\def\nh{$N_{H}$}

\def\rchisq{$\chi_{\nu}^{2}$}
\def\ergs{erg\,s$^{-1}$}
\def\ergscm2{erg\,s$^{-1}$cm$^{-2}$}
\def\ergsscm2{erg\,s$^{-2}$cm$^{-2}$}
\def\cm2{cm$^{-2}$}

\def\ee{1E\,1048.1--5937\,}

\def\xte{XTE\,J1810--197\,}

\def\1e{1E\,1547.0-5408\,}
\def\wes{CXO\,J167410.2-455216\,}
\def\sgra{SGR\,1900+14\,}
\def\sgrb{SGR\,1806--20\,}
\def\sgrc{SGR\,0526--66\,}
\def\sgrd{SGR\,1627--41\,}

\def\sgrnew{SGR\,0501+4516\,}

\title[The first outburst of the new SGR\,0501+4516]{The first outburst of the new magnetar candidate SGR\,0501+4516}
\author[N. Rea, et al.]{N. Rea$^{1}$\thanks{E-mail: n.rea@uva.nl},  G. L. Israel$^{2}$, R. Turolla$^{3,4}$, P. Esposito$^{5,6}$,
S. Mereghetti$^{5}$, D. G\"otz$^{7}$, \newauthor S. Zane$^{4}$,  A. Tiengo$^{5}$, K. Hurley$^{8}$, M. Feroci$^{9}$, M. Still$^{4}$,
V. Yershov$^{4}$,  C. Winkler$^{10}$, \newauthor R. Perna$^{11}$, F. Bernardini$^{2}$, P. Ubertini$^{8}$, L. Stella$^{2}$, S. Campana$^{12}$,  M. van der Klis$^{1}$,  \newauthor P. Woods$^{13}$ \\
$^{1}$ Astronomical Institute ``Anton Pannekoek'', University of Amsterdam, Kruislaan 403, 1098SJ, Amsterdam, The Netherlands \\
$^{2}$ INAF -- Osservatorio Astronomico di Roma, via Frascati 33, 00040, Monte Porzio Catone (RM), Italy  \\
$^{3}$ Universit\`a di Padova, Dipartimento di Fisica, via Marzolo 8, I-35131 Padova, Italy \\
$^{4}$ Mullard Space Science Laboratory, University College London, Holmbury St. Mary, Dorking, Surrey, RH5 6NT, UK \\
$^{5}$ INAF -- Istituto di Astrofisica Spaziale e Fisica Cosmica, via E.~Bassini 15, I-20133, Milano, Italy \\
$^{6}$ INFN -- Istituto Nazionale di Fisica Nucleare, Sezione di Pavia, via A.~Bassi 6, 27100 Pavia, Italy \\
$^{7}$ CEA Saclay, DSM/Irfu/Service d'Astrophysique, Orme des Merisiers, B\^at. 709, 91191 Gif sur Yvette, France \\
$^{8}$ University of California, Space Sciences Laboratory, 7 Gauss Way, 94720-7450 Berkeley, USA \\
$^{9}$ INAF -- Istituto di Astrofisica Spaziale e Fisica Cosmica, via Fosso del Cavaliere 100, I-00133 Rome, Italy \\
$^{10}$ Astrophysics Division, Research and Scientific Support Department, ESA-ESTEC, Keplerlaan 1, 2201 AZ Noordwijk, The Netherlands \\
$^{11}$ JILA, University of Colorado, Boulder, CO 80309-0440, USA \\
$^{12}$ INAF -- Osservatorio Astronomico di Brera, Via Bianchi 46, I-23807 Merate (Lc), Italy \\
$^{13}$ Dynetics, Inc., 1000 Explorer Boulevard, Huntsville, AL 35806, USA }

\input psfig.sty

\begin{document}

\pagerange{\pageref{firstpage}--\pageref{lastpage}} \pubyear{2006}

\maketitle

\label{firstpage}

\begin{abstract}

  We report here on the outburst onset and evolution of the new Soft
  Gamma Repeater SGR\,0501+4516. We monitored the new SGR with \XMM\,
  starting on 2008 August 23, one day after the source became
  burst-active, and continuing with 4 more observations in the
  following month, with the last one on 2008 September 30. Combining
  the data with the \swift-XRT and \su\, data we modelled the outburst
  decay over a three months period, and we found that the source flux
  decreased exponentially with a timescale of $t_c=23.8$\,days. In the
  first \XMM\, observation a large number of short X-ray bursts were
  observed, the rate of which decayed drastically in the following
  observations. We found large changes in the spectral and timing
  behavior of the source during the first month of the outburst decay,
  with softening emission as the flux decayed, and the non-thermal
  soft X-ray spectral component fading faster than the thermal
  one. Almost simultaneously to our second and fourth \XMM\,
  observations (on 2008 August 29 and September 2), we observed the
  source in the hard X-ray range with \INT, which clearly detected the
  source up to $\sim$100keV in the first pointing, while giving only
  upper limits during the second pointing, discovering a variable hard
  X-ray component fading in less than 10\,days after the bursting
  activation. We performed a phase-coherent X-ray timing analysis over
  about 160\,days starting with the burst activation and found
  evidence of a strong second derivative period component ($\ddot{P}$
  = -1.6(4)$\times$ 10$^{-19}$\,s~s$^{-2}$). Thanks to the
  phase-connection, we were able to study the the phase-resolved
  spectral evolution of \sgrnew\, in great detail. We also report on
  the \R\, quiescent source data, taken back in 1992 when the source
  exhibits a flux $\sim$80 times lower than that measured during the
  outburst, and a rather soft, thermal spectrum.

\end{abstract}


%%%%%%%%%%%%%%%%%%%%%%%%%%%%%%%%%%%%%%%%%%%%%
\begin{center}
\begin{figure*}
\psfig{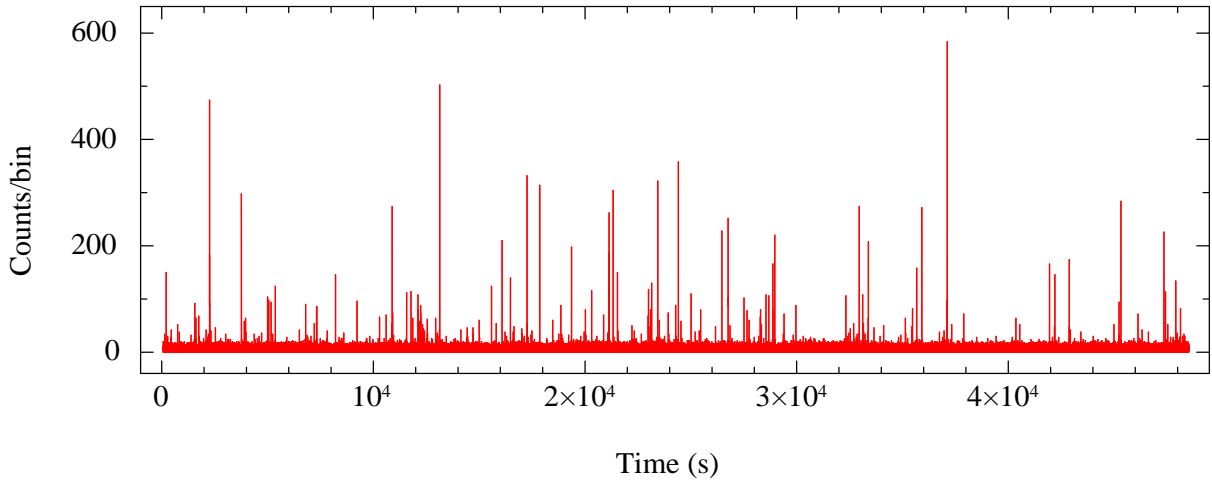}
\caption{EPIC-pn lightcurve (binned at 0.5\,s) of the 2008 August 23rd
  observation. Times are in seconds from: MJD 54701 01:07:32 (UT).}
\label{figbursts}
\end{figure*}
\end{center}
%%%%%%%%%%%%%%%%%%%%%%%%%%%%%%%%%%%%%%%%%%%%%%%%

\section{Introduction}

Over the last few years, a number of observational discoveries have
placed ``magnetars'' (ultra-magnetized isolated neutron stars) in the
limelight again. These extreme objects comprise the Anomalous X-ray
Pulsars (AXPs; 10 objects), and the Soft Gamma-ray Repeaters (SGRs; 4
objects), which are observationally very similar classes in many
respects (for a recent review see Mereghetti et al. 2008). They are
all slow X-ray pulsars with spin periods clustered in a narrow range
($P\sim$ 2--12\,s), relatively large period derivatives ($\dot P \sim
10^{-13}-10^{-10}$s\,s$^{-1}$), spin-down ages of $10^3-10^4$\,yr, and
magnetic fields, as inferred from the classical magnetic dipole
spin-down formula, of $10^{14}-10^{15}$\,G, much higher than the
electron quantum critical field
($B_{cr}\simeq4.4\times10^{13}$\,G). About a dozen AXPs and SGRs are
strong persistent X-ray emitters, with X-ray luminosities of about
$10^{34}-10^{36}$\ergs , and a few transient ones have been discovered
in recent years. A peculiarity of these neutron stars is that their
X-ray energy output is much larger than their rotational energy
losses, so they can not be only rotationally powered. Furthermore,
they lack a companion, so they can not be accretion-powered either.
Rather, the powering mechanism of AXPs and SGRs is believed to reside
in the neutron star ultra-strong magnetic field (Duncan \& Thompson
1992; Thompson \& Duncan 1993). Other scenarios, beside the
``magnetar'' model, were proposed to explain AXP and SGR emission,
such as the fossil disk (Chatterjee, Hernquist \& Narayan 2000; Perna,
Hernquist, \& Narayan 2000) and the quark-star model (Ouyed, Leahy, \&
Niebergal 2007a,b).

In the 0.1--10\,keV energy band, magnetars spectra are relatively soft
and empirically modeled by an absorbed blackbody
($kT\sim$\,0.2--0.6~keV) plus a power-law ($\Gamma\sim$\,2--4).
Thanks to \INT--ISGRI and \RXTE--HEXTE, hard X-ray emission up to
$\sim$200\,keV has recently been detected from some sources (Kuiper et
al. 2004, 2006; Mereghetti et al. 2005; G\"otz et al. 2006).  This
discovery has opened a new window on magnetars studies and has shown
that their energy output may be dominated by hard, rather than soft
emission.

At variance with other isolated neutron stars, AXPs and SGRs exhibit
spectacular episodes of bursting and flaring activity, during which
their luminosity may change up to 10 orders of magnitude on timescales
down to few milliseconds. Different types of X-ray flux variability
have been observed, ranging from slow and moderate flux changes up to
a factor of a few on timescales of years (shown by virtually all
members of the class), to more intense outbursts with flux variations
up to $\sim$100 lasting for $\sim$1-3 years, and to short and intense
X-ray burst activity on sub-second timescales (see Kaspi 2007 and
Mereghetti 2008 for reviews of X-ray variability).

%%%%%%%%%%%%%%%%%%%%%%%%%%%%%%%%%%%%%%%%%%%%%%%%%%%%%%%%%%%%%%%%%%%%%%
\begin{table*}
\begin{center}
\begin{tabular}{lccccc}
\hline
\hline
 \multicolumn{1}{l}{Parameters} & \multicolumn{1}{c}{2008-08-23} &
 \multicolumn{1}{c}{2008-08-29} & \multicolumn{1}{c}{2008-08-31} &
 \multicolumn{1}{c}{2008-09-02} & \multicolumn{1}{c}{2008-09-30} \\
\hline
Start (UT)    & 01:07:36  & 07:10:28  &  12:09:45 &  10:00:38 & 02:18:44\\
End (UT) & 14:35:33 & 13:58:20  &  14:59:58  & 15:41:49 & 11:22:15 \\
Exposure (ks) &  48.9 & 24.9 &  10.2 &  20.5 & 31.0\\
Counts/s (pn) & $8.520\pm0.016$  &  $7.08\pm0.02$ &  $6.60\pm0.03$  & $6.05\pm0.02$  &  $3.23\pm0.01$\\
%Counts/s (MOS2) & $3.17\pm0.01$   &  &    &   &  \\
%Counts/s (RGS1) & $0.07\pm0.001$  &  &    &   &  \\
\hline
Pulse Period (s)   & 5.7620694(1) & 5.7620730(1) & 5.7620742(1) & 5.7620754(1)   & 5.7620917(1) \\
Pulsed Fraction (\%) &  41(1) & 35(1)  & 38(1) & 38(1) & 43(1) \\
N. bursts & 80 & 2  &  0  & 0  & 0  \\
\hline
\hline
\end{tabular}

\caption{{\em Top table}: Summary of the first 5 \XMM\, observations
  of \sgrnew. The exposure time refers to the pn camera.  Count-rates
  are background-corrected, and refers to the pn in Small Window, except
  for the last observation which was in Large Window. {\em Bottom
    table}: Timing properties of \sgrnew. The pulsed fraction is defined as
  the background-corrected $(max-min)/(max+min)$ in the 0.3-12keV
  energy band. The number of bursts refers to spikes detected at $>$35
  count/s.\label{obslog}}
\end{center}
\end{table*}

%%%%%%%%%%%%%%%%%%%%%%%%%%%%%%%%%%%%%%%%%%%%%%%%%%%%%%%%%%%%%%%%%%%%%%%%%%%%%%%%

In particular, SGRs are characterized by periods of activity during
which they emit numerous {\em short bursts} in the hard X-ray / soft
gamma-ray energy range ($t\sim0.1-0.2$\,s; $L\sim10^{38}-10^{41}$
erg/s). This is indeed the defining property that led to the discovery
of this class of sources. In addition, they have been observed to emit
{\em intermediate flares}, with typical durations of $t\sim1-60$\,s
and luminosities of $L\sim10^{41}-10^{43}$ erg/s, and spectacular {\em
  Giant Flares}.  The latter are rare and unique events in the X-ray
sky, by far the most energetic ($\sim10^{44}-10^{47}$ erg/s) Galactic
events currently known, second only to Supernova explosions. Indeed,
the idea that SGRs host an ultra-magnetized neutron star was
originally proposed to explain the very extreme properties of their
bursts and flares: in this model the frequent short bursts are
associated with small cracks in the neutron star crust, driven by
magnetic diffusion, or, alternatively, with the sudden loss of
magnetic equilibrium through the development of a tearing instability,
while the giant flares would be linked to global rearrangements of the
magnetic field in the neutron stars magnetosphere and interior
(Thompson \& Duncan 1995; Lyutikov 2003).

Bursts and flares do not seem to repeat with any regular, predictable
pattern. Giant flares have been so far observed only three times from
the whole sample of SGRs (from \sgrc\, in 1979, Mazets et al.~1979;
from \sgrb\, in 1998, Hurley et al.~1999; and from \sgra\, in 2004,
e.g. Hurley et al. 2005, Palmer et al. 2005), and never twice from the
same source.  As far as short bursts and intermediate flares are
concerned, while some SGRs (such as \sgrb) are extremely active
sources, in other cases no bursts have been detected for many years
(as in the case of SGR 1627-41, that re-activated in May 2008 after a
10-yr long stretch of quiescence; Esposito et al. 2008). This suggests
that a relatively large number of members of this class has not been
discovered yet, and may manifest themselves in the future.

On 2008 August 22, a new SGR, namely \sgrnew, was discovered (the
first in ten years), thanks to the \swift-BAT detection of a series of
short X-ray bursts and intermediate flares (Holland et al. 2008;
Barthelmy et al. 2008).  X-ray pulsations were observed by \RXTE\, at
a period of 5.7s, confirming the magnetar nature of this source
(G{\"o}{\u g}{\"u}{\c s} et al. 2008), and its counterpart was
identified in the infrared and optical bands (Tanvir et al. 2008; Rea
et al. 2008b; Fatkhullin et al. 2008; Rol et al. 2008). Prompt radio
observations to search for the on-set of radio pulsation and of a
persistent counterpart failed to reveal any emission in this band in
the first days after the outburst activation (Hessels et al. 2008;
Kulkarni \& Frail 2008b; Gelfand et al. 2008).

In this paper, we present a series of 5 \xmm\ observations of \sgrnew;
the first one was performed only 1 day after the SGR activation, and
the last one after 38 days. We also report on two \INT\, observations;
the first was performed almost simultaneously with the second \XMM\,
observation, while the other one was performed soon after the fourth
\XMM\, pointing. We used the \swift-XRT monitoring to model the
outburst decay and the spin period evolution of the source until
$\sim$160\,days after the onset of the bursting activity.  We also
report on the 1992 \R\, observation of its quiescent counterpart.  We
present details of the observation and analysis in \S\,\ref{obs}, and
results in \S\,\ref{timing} and \ref{spectra}. Discussion follows in
\S\,\ref{discussion}.

%%%%%%%%%%%%%%%%%%%%%%%%%%%%%%%%%%%%%%%%%%%%%
\begin{center}
\begin{figure}
\centerline{\psfig{figure=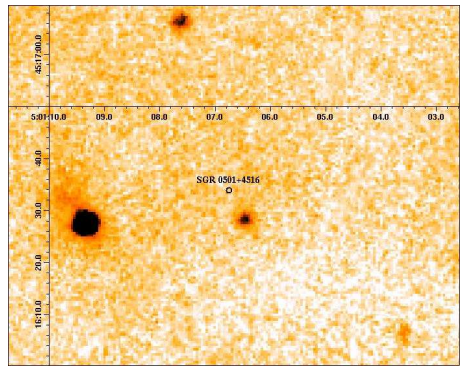,width=6cm}}
\caption{Co-added image of all the OM observations in the UVW1 filter.
The four bright objects are USNO B1 stars.}
\label{figom}
\end{figure}
\end{center}
%%%%%%%%%%%%%%%%%%%%%%%%%%%%%%%%%%%%%%%%%%%%%%%%%%%%%%%%%%%%%%%%

%%%%%%%%%%%%%%%%%%%%%%%%%%%%%%%%%%%%%%%%%%%%%

\begin{figure*}
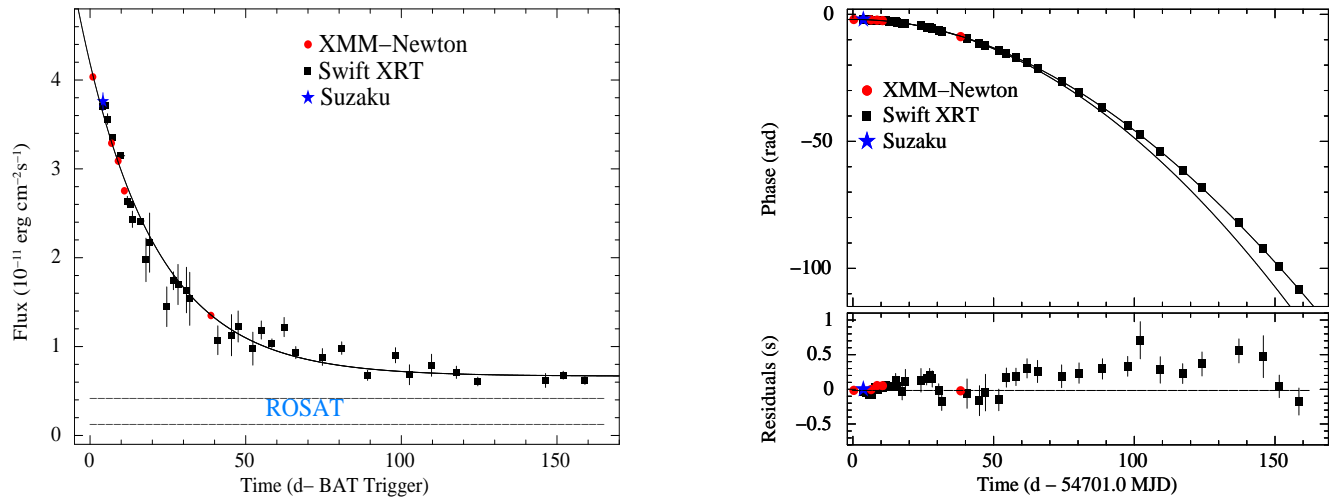

\begin{center}
\hbox{
\psfig{figure=sgr0501_fluxdecay_last.ps,width=10cm,height=6.5cm,angle=-90}
\hspace{-0.2cm}
\psfig{figure=timingSGR0501_last.ps,angle=-90,height=6.5cm,width=9cm}}

\caption{{\em Left hand panel}: the outburst decay of the persistent
  X-ray flux of \sgrnew\, fitted with an exponential function (see
  \S\ref{discussion} for details). We refer here as BAT trigger: MJD
  54700.0 12:41:59.000 (UT). The fluxes are absorbed and in the
  1--10\,keV energy range for \XMM, \swift, and \su\, (Enoto et
  al. 2009), while the \R\, flux is extrapolated to the same band, and
  refers to two different spectral models (see \S\ref{spectra} for
  details).{\em Right hand panel}: The 0.5--10\,keV pulse phase
  evolution with time, together with the time residuals with respect
  to the phase coherent timing solution discussed in the text and
  including${\rm P}$/$\dot{\rm P}$/$\ddot{\rm P}$ components. The
  solid lines in the upper panel represent the timing solution with
  (top line) and without (low line) the cubic term.}

\label{figfluxdecay}
\label{phasecorr}
\label{phase}
\end{center}
\end{figure*}
%%%%%%%%%%%%%%%%%%%%%%%%%%%%%%%%%%%%%%%%%%%%%%%%%%%%%%%%%%%%%%%%

\section{Observations and analysis}
\label{obs}

\subsection{XMM-Newton}
\label{xmm}

The \xmm\, Observatory (Jansen et al. 2001) observed \sgrnew\, on
August/September 2008 (see Tab.\ref{obslog}) with the EPIC instruments
(pn and MOSs; Turner et al. 2001; Str\"uder et al. 2001), the
Reflecting Grating Spectrometer (RGS; den Herder et al. 2001), and the
Optical Monitor (OM; Mason et al. 2001).

Data were processed using SAS version 7.1.0 with the most up to date
calibration files (CCF) available at the time the reduction was
performed (October 2008). Standard data screening criteria were
applied in the extraction of scientific products. Soft proton flares
were not observed in any of the observations, resulting in the total
on-source exposure times listed in Tab\,\ref{obslog}.

\subsubsection{EPIC and RGS}

For four of the observations the pn camera was set in {\tt Small
  Window} mode in order to reduce pile--up, while for the 2008
September 30th observation it was in {\tt Large Window} mode. The MOS1
camera was in {\tt Full Frame} for the first observation, and in {\tt
  Small Window} for all the other pointings. On the other hand, the
MOS2 was in {\tt Timing} mode, except for the last observation where
it was set in {\tt Small Window} mode. All other MOS CCDs were in {\tt
  Prime Full Window} mode. Thick filters were used for all the
instruments, and pile-up was present only in the first MOS1
observation, which we ignored in the rest of the analysis. No
transients were present in any imaging camera, so we are confident
that the MOS2 in non-imaging mode did not collect photons from
anything else than our target.

We performed a 2-dimensional or 1-dimensional PSF fitting, for the
data obtained with the EPIC cameras in imaging mode or timing mode,
respectively. The extraction radius was chosen in such a way as to
obtain more than 90\% of the source counts.

We then extracted the source photons, for the cameras set-up in
imaging mode, from a circular region with 30\arcsec radius, centered
at the source position (RA 05:01:06.607, Dec $+$45:16:33.47 at J2000,
with a 1$\sigma$ error of 1\farcs5 which refers to the absolute
astrometric \XMM\, accuracy (Kirsh et al. 2004))\footnote{Consistent
  with the more accurate \CXO\, determination: RA 05:01:06.756, Dec
  $+$45:16:33.92 (0.11\arcsec error circle; Woods et al. 2008)}. The
background was obtained from a similar region as far away as possible
from the source location in the same CCD. For the MOS2 camera in
timing mode we extracted the photons from RAWX 274-334, and a similar
region was used for the background extraction, although as far as
possible from the source position. Only photons with PATTERN$\leq 4$
were used for the pn, with PATTERN$\leq 12$ for the MOS2 when in
imaging mode, and with PATTERN$=0$ were used for MOS2 observations in
timing mode. All the photon arrival times have been corrected to refer
to the barycenter of the Solar System.

Thanks to the high timing and spectral resolution\footnote{see
  http://xmm.esac.esa.int/ for details.} of the pn and MOS cameras,
and to the high spectroscopic accuracy of the RGS, we were able to
perform timing and spectral analysis, as well as pulse phase
spectroscopy. Both the MOSs and pn cameras gave consistent timing and
spectral results, and we report only on the pn results (see
Tab.\,\ref{obslog} for the pn source count rates for all five
observations), and the RGS is used only to constrain the presence of
narrow lines (see \S\,\ref{spectra}).

For the timing (\S\ref{timing}) and spectral analysis
(\S\ref{spectra}) we removed the bursts observed in the first two
observations (August 23rd and 29th) discarting all the photons
corresponding to intervals where the source count rate exceeded 35
counts\,s$^{-1}$ (a detailed analysis of the bursts themselves will be
reported elsewhere).

\subsubsection{Optical Monitor}
\label{obsom}

Twenty five OM images of the field were obtained simultaneously to the
X-ray observations through the UVW1 lenticular filter. One further
image was obtained through the U filter. The UVW1 has an effective
transmission range of $\lambda=$2410--3565\,A, peak efficiency at
$\lambda$2675\,A, full-width half-maximum image resolution of 2\arcsec
and a Vega-spectrum zeropoint of $m$=17.20. The U has an effective
transmission range of $\lambda=$3030--3890\,A, peak efficiency at
$\lambda=$3275\,A, full-width half-maximum image resolution of
1.55\arcsec\, and a Vega-spectrum zeropoint of $m$=18.26. Modulo-8
fixed photon pattern and scattered background light were removed from
individual images before correcting optical distortion and converting
images to J2000 celestial coordinates. The \XMM\, star trackers
provide absolute pointing accurate to 1\farcs8. To refine astrometry,
a correction is performed to individual images by cross-correlating
source positions in the OM with counterparts in the USNO-B1.0
catalogue (Monet 2003). The UVW1 images were mosaicked to produce a
70\,ks summed exposure. The U band image was accumulated over an
exposure time of 4\,ks. Aperture photometry was performed on the
source position of \sgrnew\, using a standard 17\farcs5 radius
circular aperture for the UVW1 image and 3\arcsec\, for the U image,
consistent with the calibrated zeropoint.

%%%%%%%%%%%%%%%%%%%%%%%%%%%%%%%%%%%%%%%%%%%%%%%%%%%%%%%%%%%%%%%%%

\begin{center}
\begin{figure}
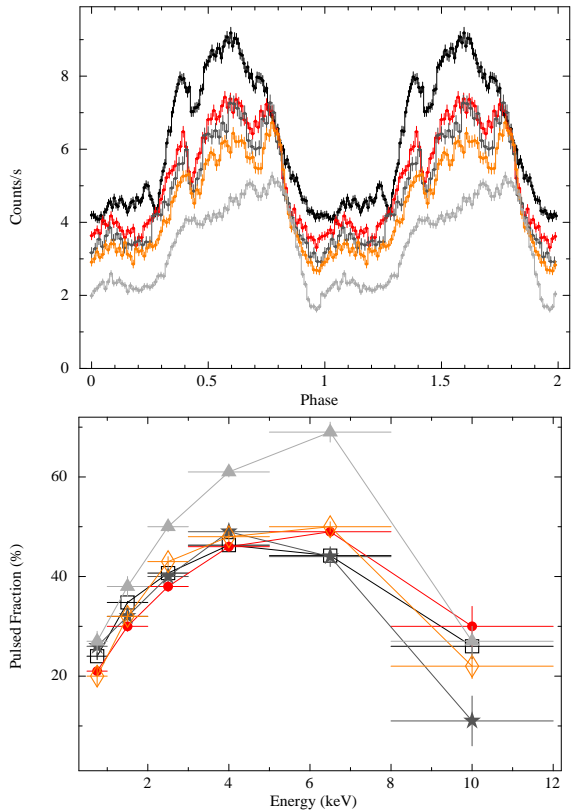

\vbox{
\psfig{figure=efold_total_all_new3.ps,width=7.5cm,angle=-90}
\vspace{0.1cm}
\psfig{figure=PFvsE_all_1sin_new.ps,width=7.5cm,angle=-90}}
\caption{{\em Top panel}: Pulse profiles of the 5 \XMM\, observations
in the 0.3-12\,keV energy band. {\em Bottom panel} Pulsed fraction
dependence with energy for the same observations. In both panels
the black, red, dark grey, orange and light grey colors refer to the
five observations ordered by increasing epoch.}
\label{pfenergy}
\end{figure}
\end{center}

%%%%%%%%%%%%%%%%%%%%%%%%%%%%%%%%%%%%%%%%%%%%%%%%%%%%%%%%%%%%%%%%%%%%%%

No XMM-OM source is detected within this aperture to 3$\sigma$
magnitude upper limits of $m_{\rm U}>$ 22.1 and $m_{\rm UVW1}>$ 23.7
(see Fig. \ref{figom}). We also searched for possible counterparts to
the X-ray bursts in the XMM-OM exposures in the UVW1 filter during the
first \XMM\, observation. We did not find any signature for such
bursts in the UVW1 filter with a 3$\sigma$ upper limit on each 4\,ks
image of $m_{\rm UVW1}>$ 22.05 .

\subsection{INTEGRAL}
\label{integral}

\INT\ (Winkler et al. 2003) observed \sgrnew\ twice, soon after its
discovery: the first observation (orbit 717), soon after its
discovery, started on 2008 August 27 at 00:31 (UT) as a ToO
observation (ending on August 28th 08:36 UT), and the second
observation in the framework of the Core Programme observations of the
Perseus Arm region starting on 2008 September 5 at 05:48, and ending
at 07:40 (UT) on September 10th (orbits 720 and 721). We analyzed the
IBIS/ISGRI data of both observations. IBIS (Ubertini et al. 2003) is a coded
mask telescope with a wide (29$^{\circ}\times$29$^{\circ}$) field of
view, sensitive in the 15 keV--10 MeV energy range. We restricted our
analysis to the ISGRI (Lebrun et al. 2003) data, taken by the IBIS low energy
(15\,keV--1\,MeV) CdTe detector layer, since ISGRI the most sensitive
instrument on board \INT\ at energies $<$ 300\,keV.

For the first observation an effective exposure of 204\,ks was
accumulated at the source position. During this observation, the
source was still burst-active and indeed at least 4 weak bursts were
detected in the ISGRI data (Hurley et al. 2008). In the 18--60\,keV
image the source is detected at a $\sim$4.2\,$\sigma$ confidence
level, corresponding to a count rate of 0.31$\pm$0.08 counts s$^{-1}$,
while in the 60--100 keV band the source was detected at a $\sim$3.5
sigma level (0.25$\pm$0.07 counts\,s$^{-1}$). Above 100 keV the source
is not detected and the 3$\sigma$ upper limit is 0.2 counts\,s$^{-1}$
(100--200\,keV).  The ISGRI response matrices were rebinned to match
the above two channels and the detected flux values were used in the
broad band spectral analysis (see below \S\ref{spectra}).

We performed the same analysis on the Core Programme data. In this
case the exposure time was 361\,ks at the position of the source. No
persistent or burst emission was detected in this second
observation. We could infer a 3$\sigma$ upper limit in the 18--60 keV
energy band of 0.18 counts\,s$^{-1}$, implying a decrease of the hard
X-ray flux in about 10\,days of a factor of $\sim$2.

%%%%%%%%%%%%%%%%%%%%% efolds  %%%%%%%%%%%%%%%%%%%%%%%%%
\begin{center}
\begin{figure*}
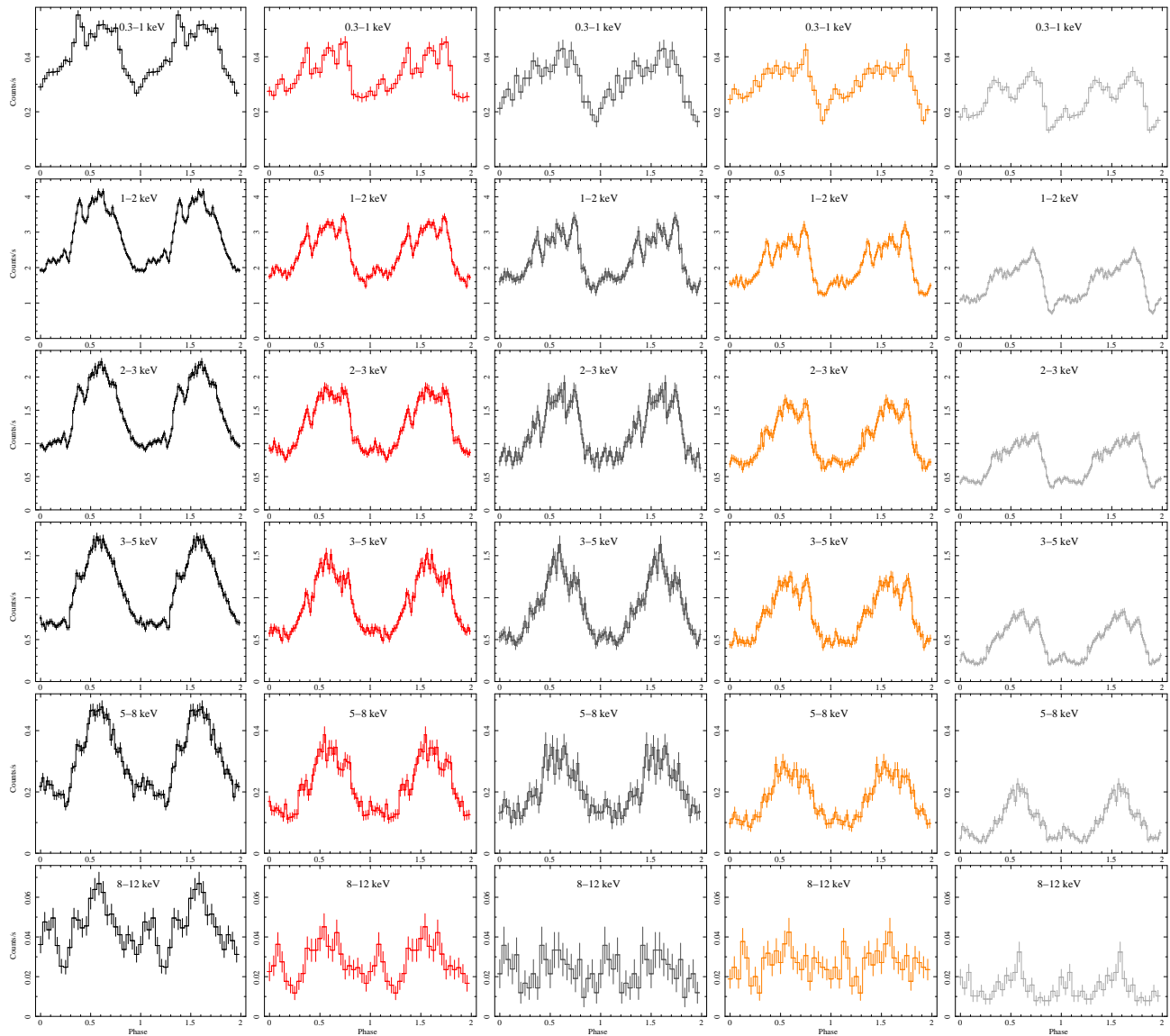

%\hspace{0.1cm}
\hbox{
\vbox{
\psfig{figure=efold_2308_03-1new4.ps,width=3.5cm,height=2.5cm,angle=-90}
\psfig{figure=efold_2308_1-2new4.ps,width=3.5cm,height=2.5cm,angle=-90}
\psfig{figure=efold_2308_2-3new4.ps,width=3.5cm,height=2.5cm,angle=-90}
\psfig{figure=efold_2308_3-5new4.ps,width=3.5cm,height=2.5cm,angle=-90}
\psfig{figure=efold_2308_5-8new4.ps,width=3.5cm,height=2.5cm,angle=-90}
\psfig{figure=efold_2308_8-12new4.ps,width=3.5cm,height=2.5cm,angle=-90}}
\vbox{
\psfig{figure=efold_2908_03-1new4.ps,width=3.3cm,height=2.5cm,angle=-90}
\psfig{figure=efold_2908_1-2new4.ps,width=3.3cm,height=2.5cm,angle=-90}
\psfig{figure=efold_2908_2-3new4.ps,width=3.3cm,height=2.5cm,angle=-90}
\psfig{figure=efold_2908_3-5new4.ps,width=3.3cm,height=2.5cm,angle=-90}
\psfig{figure=efold_2908_5-8new4.ps,width=3.3cm,height=2.5cm,angle=-90}
\psfig{figure=efold_2908_8-12new4.ps,width=3.3cm,height=2.5cm,angle=-90}}
\vbox{
\psfig{figure=efold_3108_03-1new4.ps,width=3.3cm,height=2.5cm,angle=-90}
\psfig{figure=efold_3108_1-2new4.ps,width=3.3cm,height=2.5cm,angle=-90}
\psfig{figure=efold_3108_2-3new4.ps,width=3.3cm,height=2.5cm,angle=-90}
\psfig{figure=efold_3108_3-5new4.ps,width=3.3cm,height=2.5cm,angle=-90}
\psfig{figure=efold_3108_5-8new4.ps,width=3.3cm,height=2.5cm,angle=-90}
\psfig{figure=efold_3108_8-12new4.ps,width=3.3cm,height=2.5cm,angle=-90}}
\vbox{
\psfig{figure=efold_0209_03-1new4.ps,width=3.3cm,height=2.5cm,angle=-90}
\psfig{figure=efold_0209_1-2new4.ps,width=3.3cm,height=2.5cm,angle=-90}
\psfig{figure=efold_0209_2-3new4.ps,width=3.3cm,height=2.5cm,angle=-90}
\psfig{figure=efold_0209_3-5new4.ps,width=3.3cm,height=2.5cm,angle=-90}
\psfig{figure=efold_0209_5-8new4.ps,width=3.3cm,height=2.5cm,angle=-90}
\psfig{figure=efold_0209_8-12new4.ps,width=3.3cm,height=2.5cm,angle=-90}}
\vbox{
\psfig{figure=efold_3009_03-1new4.ps,width=3.3cm,height=2.5cm,angle=-90}
\psfig{figure=efold_3009_1-2new4.ps,width=3.3cm,height=2.5cm,angle=-90}
\psfig{figure=efold_3009_2-3new4.ps,width=3.3cm,height=2.5cm,angle=-90}
\psfig{figure=efold_3009_3-5new4.ps,width=3.3cm,height=2.5cm,angle=-90}
\psfig{figure=efold_3009_5-8new4.ps,width=3.3cm,height=2.5cm,angle=-90}
\psfig{figure=efold_3009_8-12new4.ps,width=3.3cm,height=2.5cm,angle=-90}}}
\caption{Pulse profiles (phase vs counts/s) as a function of energy for all five
  \XMM\, observations of \sgrnew. Each column displays one \XMM\,
  observation with epoch increasing from left to right.}
\label{pulsenergy}
\end{figure*}
\end{center}
%%%%%%%%%%%%%%%%%%%%%%%%%%%%%%%%%%%%%%%%%%%%%%%%%%%%%%%%%%%%%%%%

%%%%%%%%%%%%%%%%%%%%%%%%%%%%%%%%%%%%%%%%%%%%%%%%%%%%%%%%%%%%%%%%%

\begin{figure*}
\begin{center}
\hspace{0.1cm}
\hbox{
\vbox{
\psfig{figure=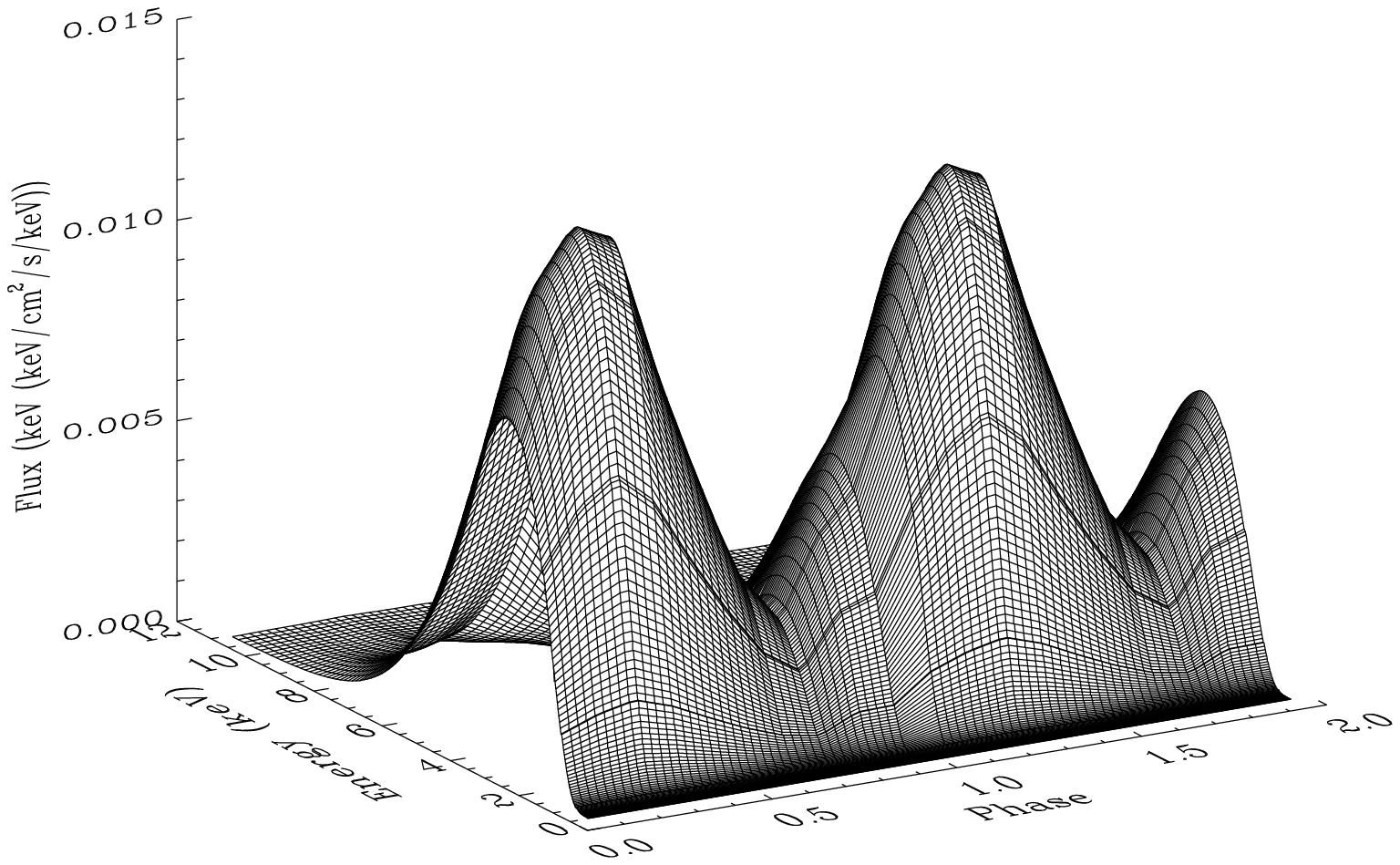,width=7cm}
\psfig{figure=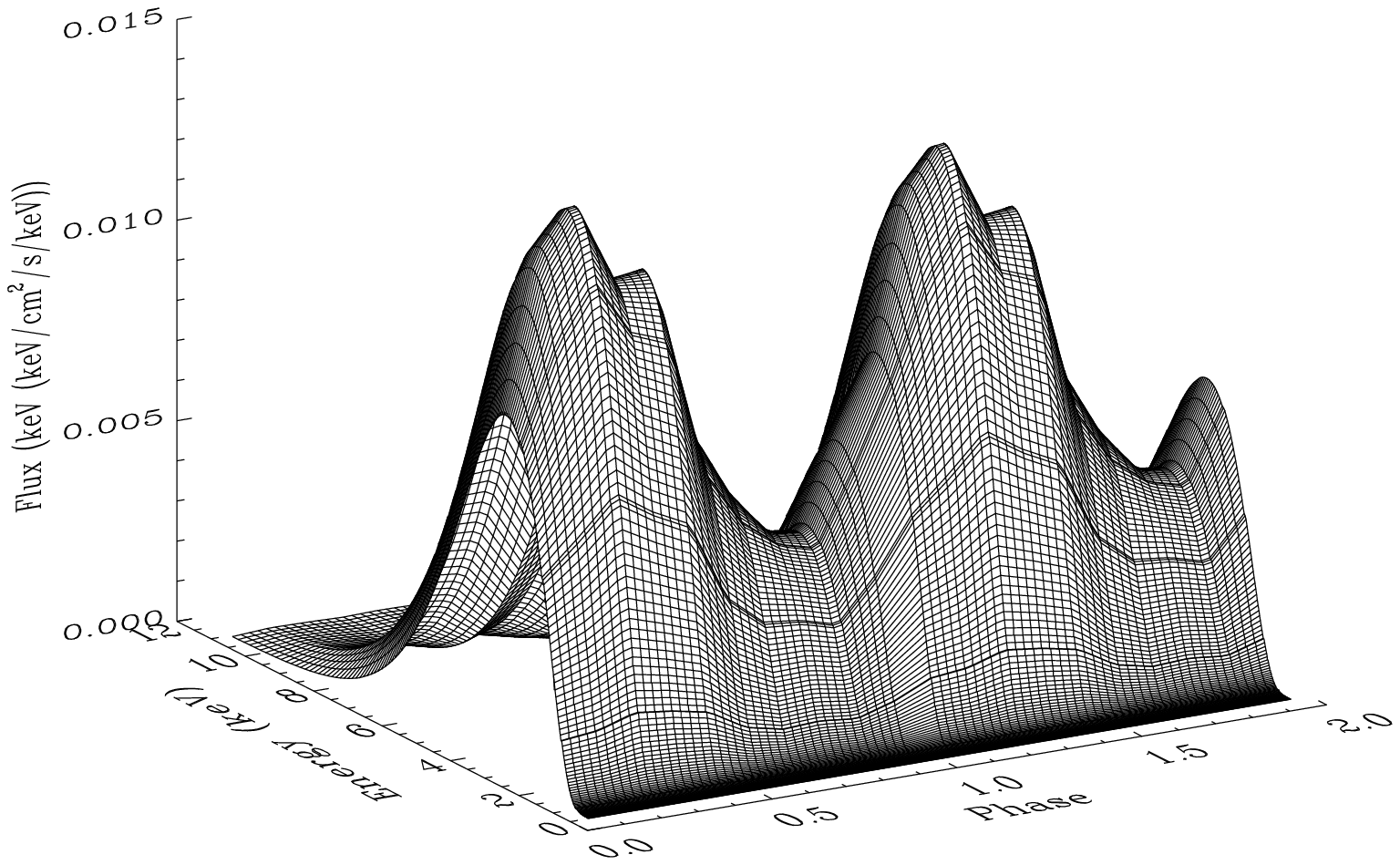,width=7cm}
\psfig{figure=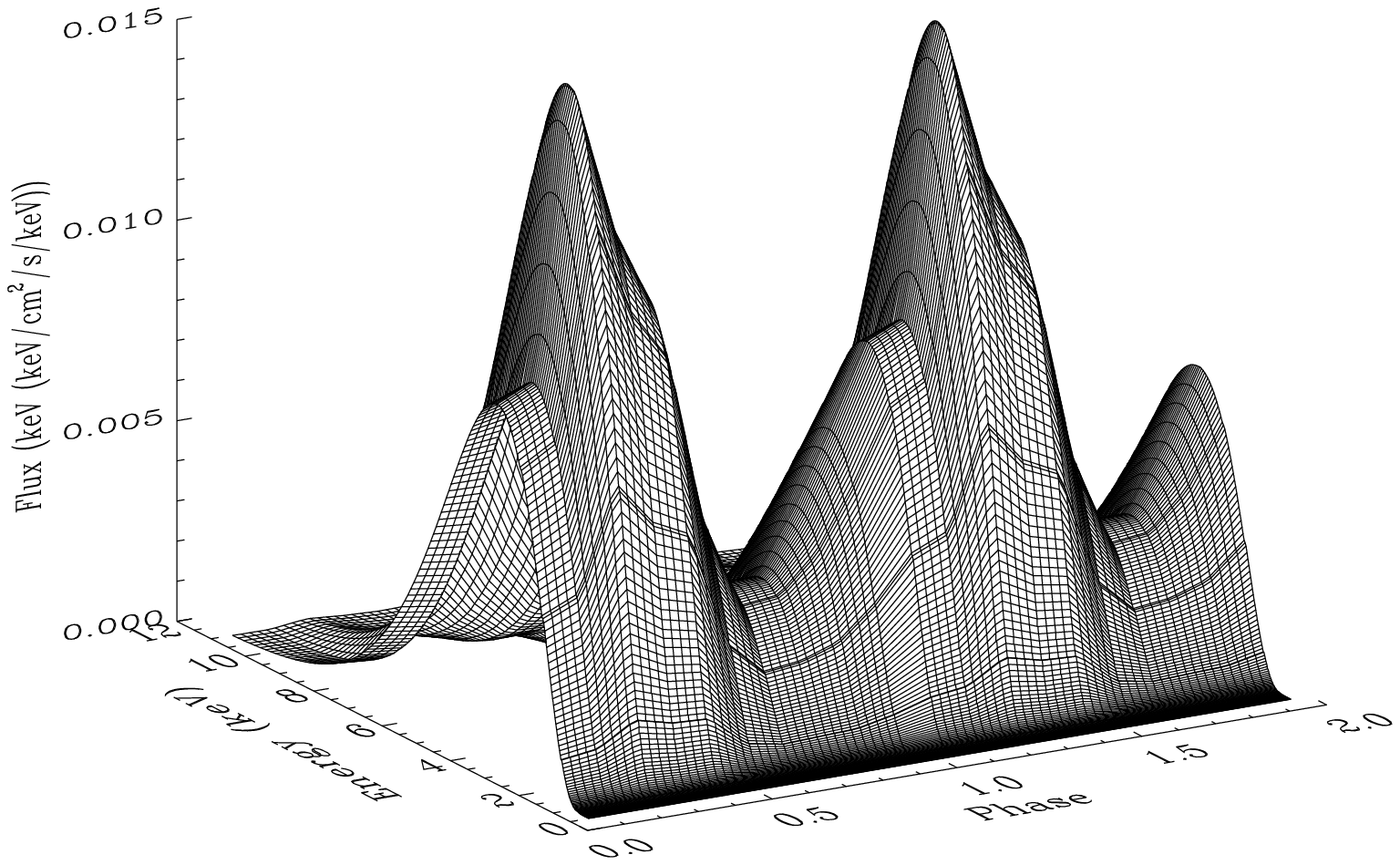,width=7cm}
\psfig{figure=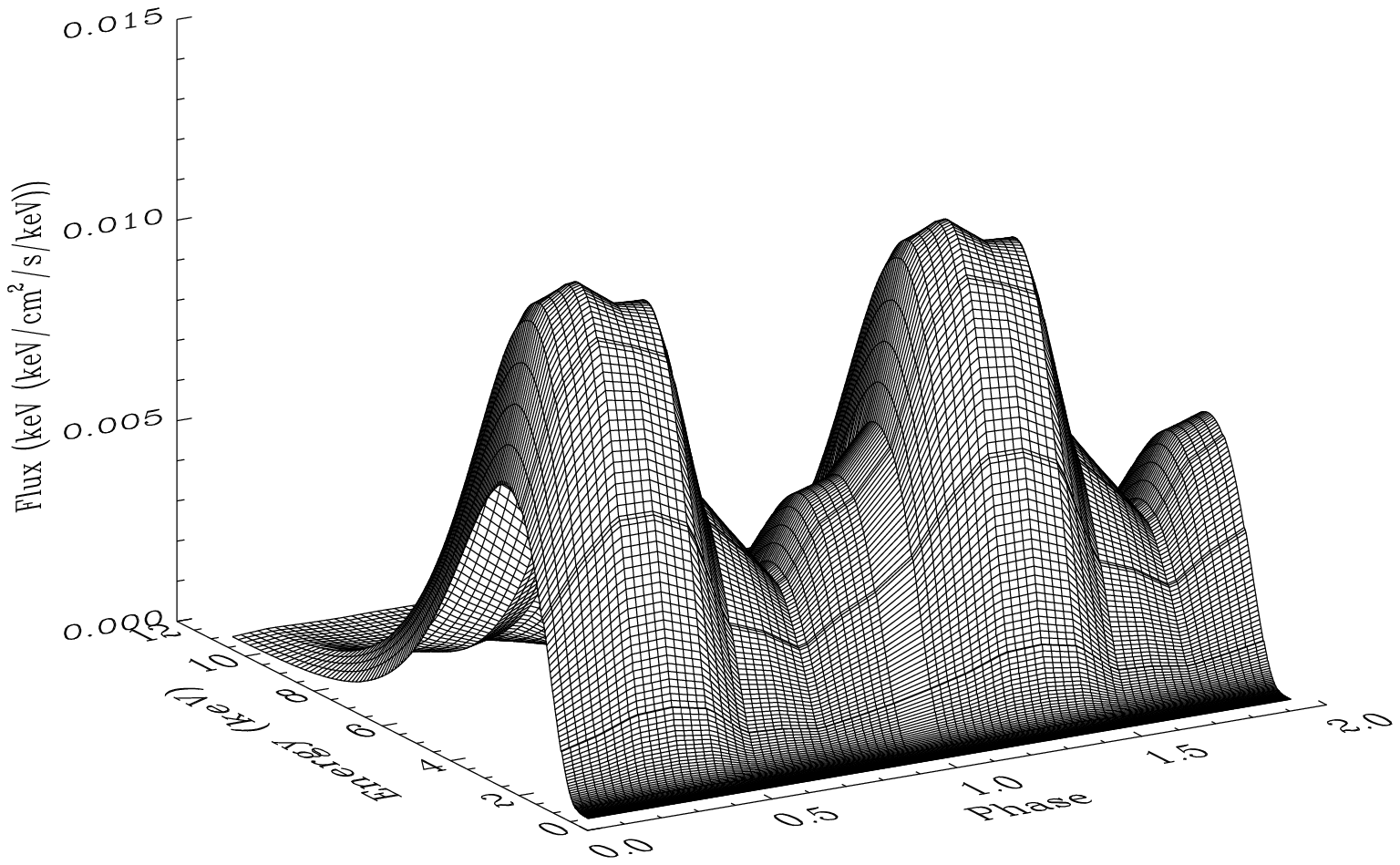,width=7cm}
\psfig{figure=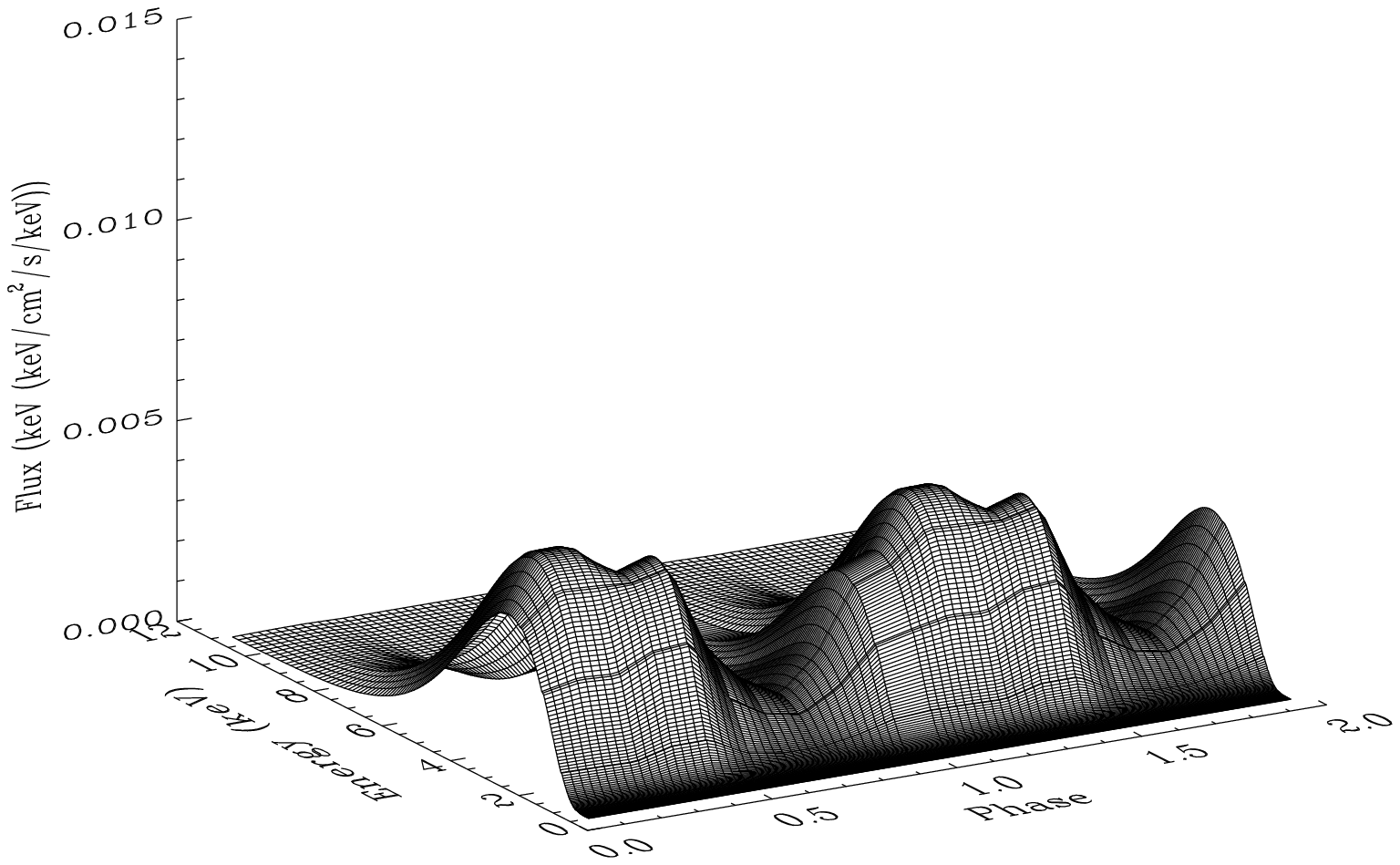,width=7cm}}
\vbox{
\psfig{figure=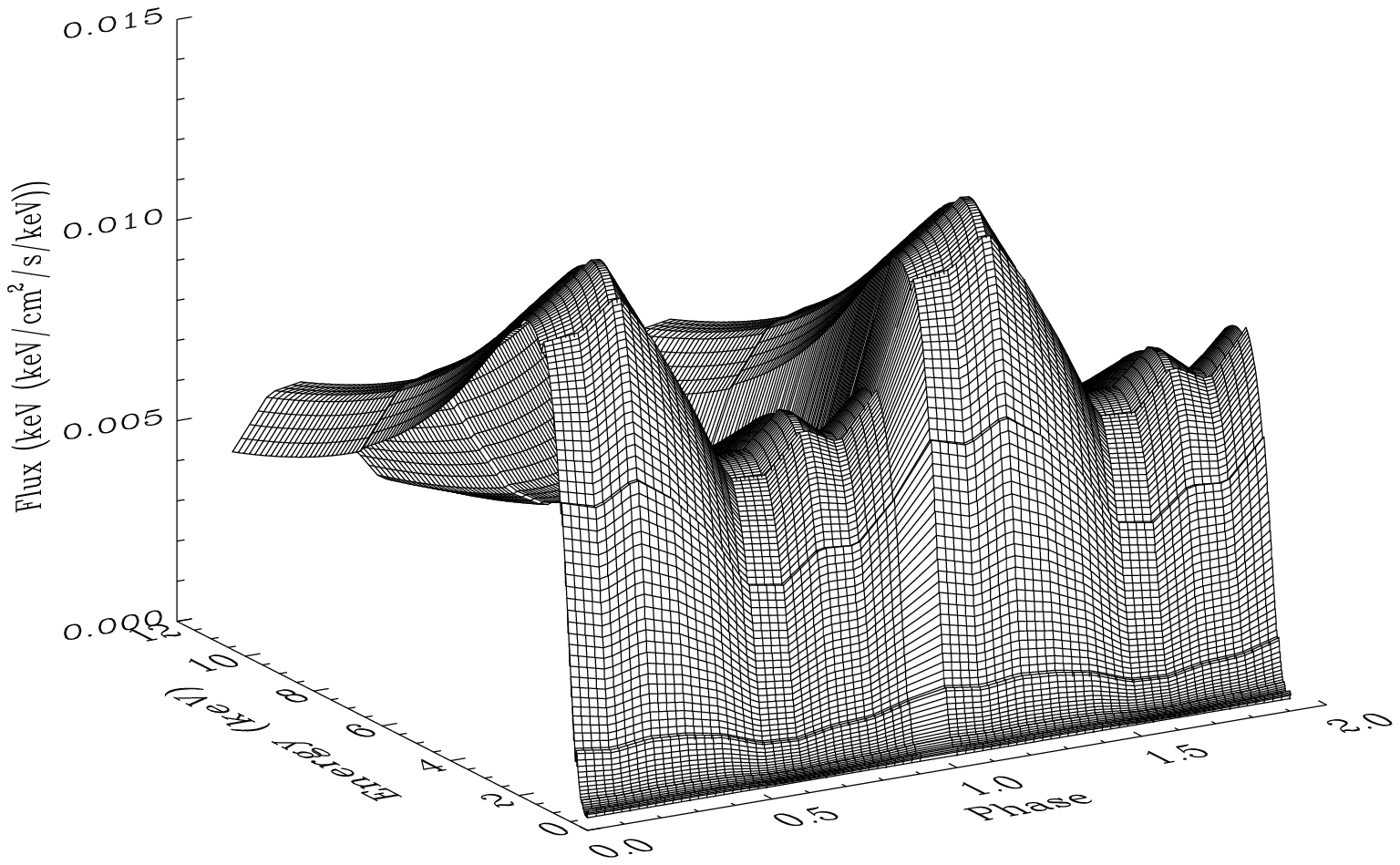,width=7cm}
\psfig{figure=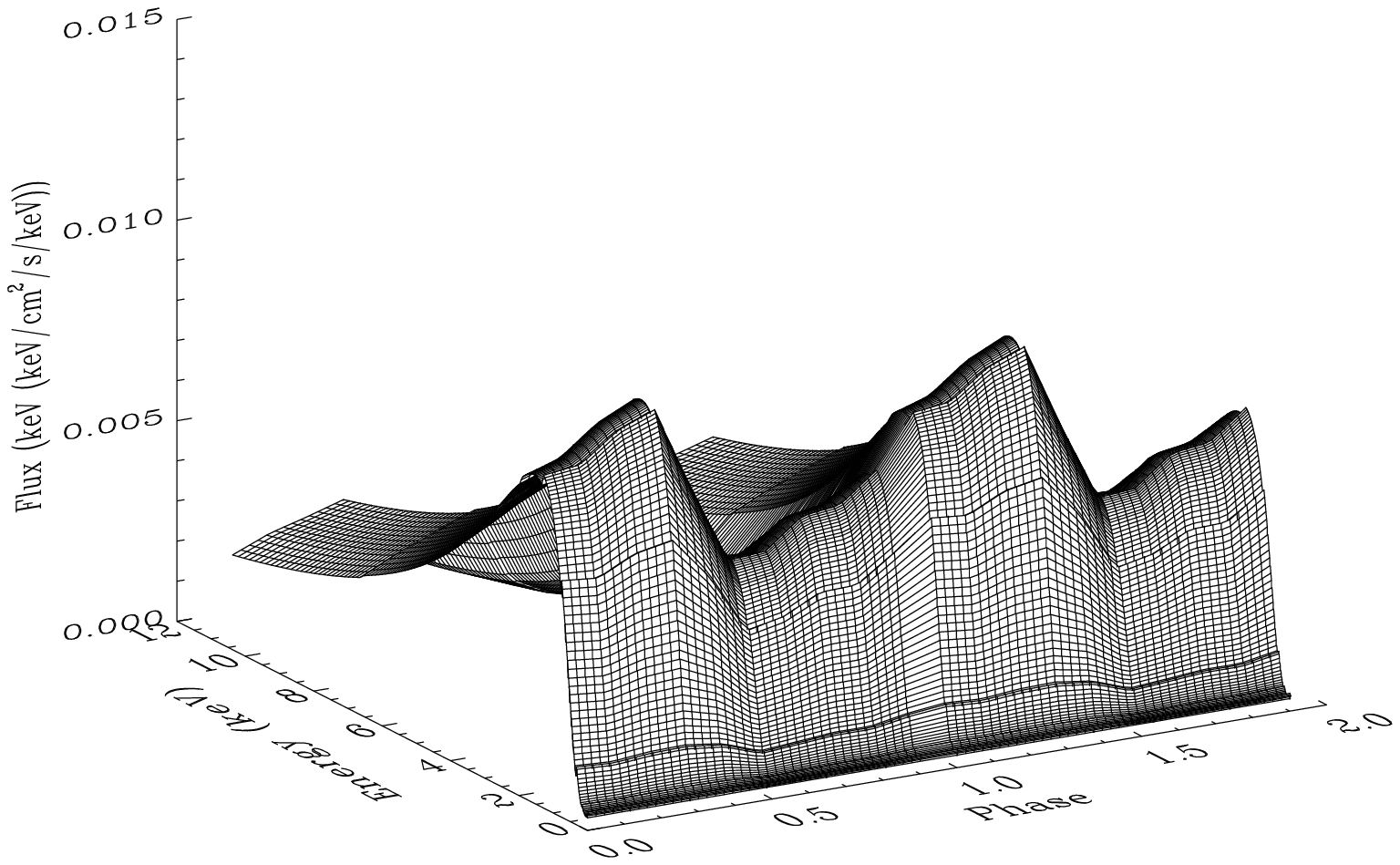,width=7cm}
\psfig{figure=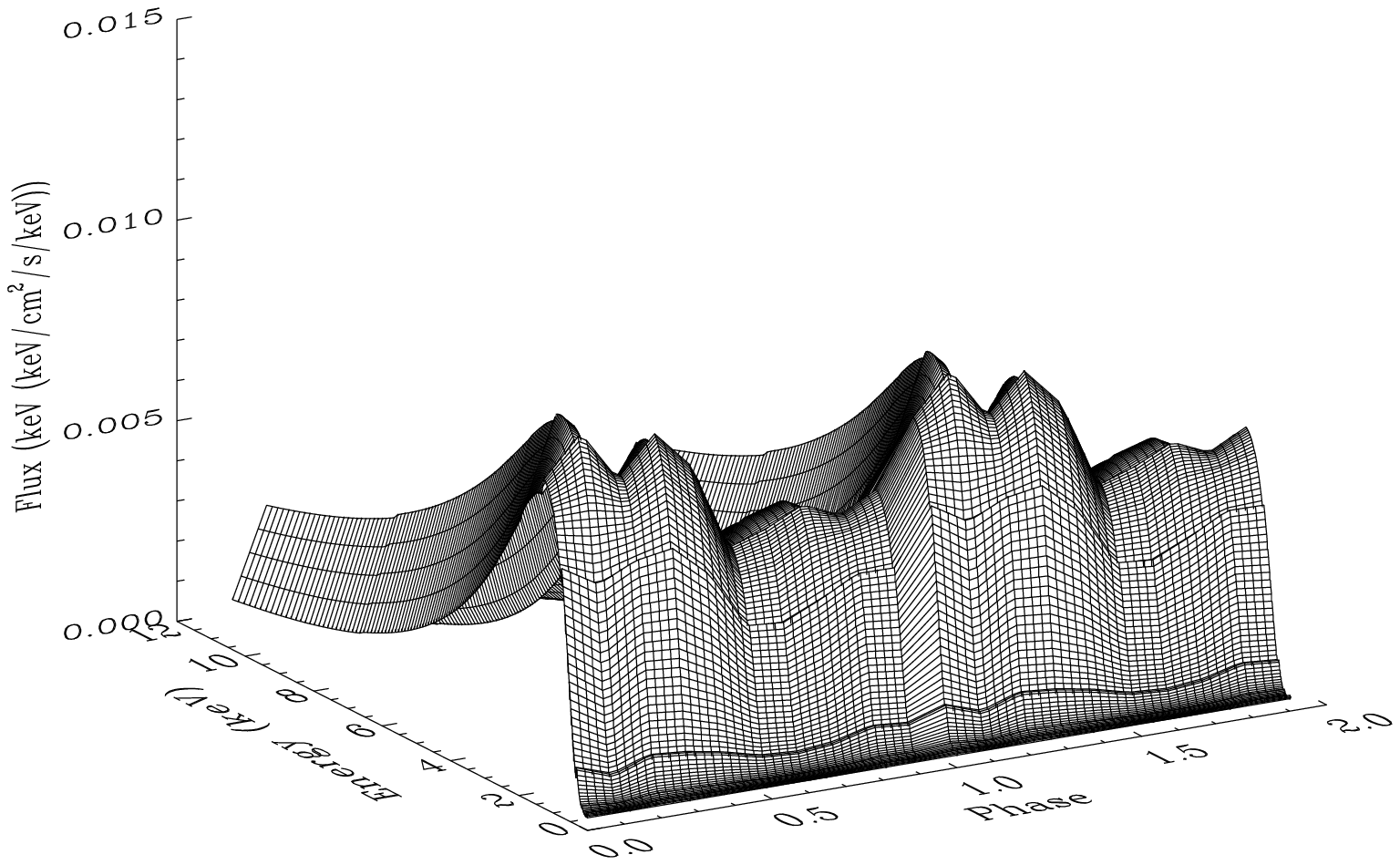,width=7cm}
\psfig{figure=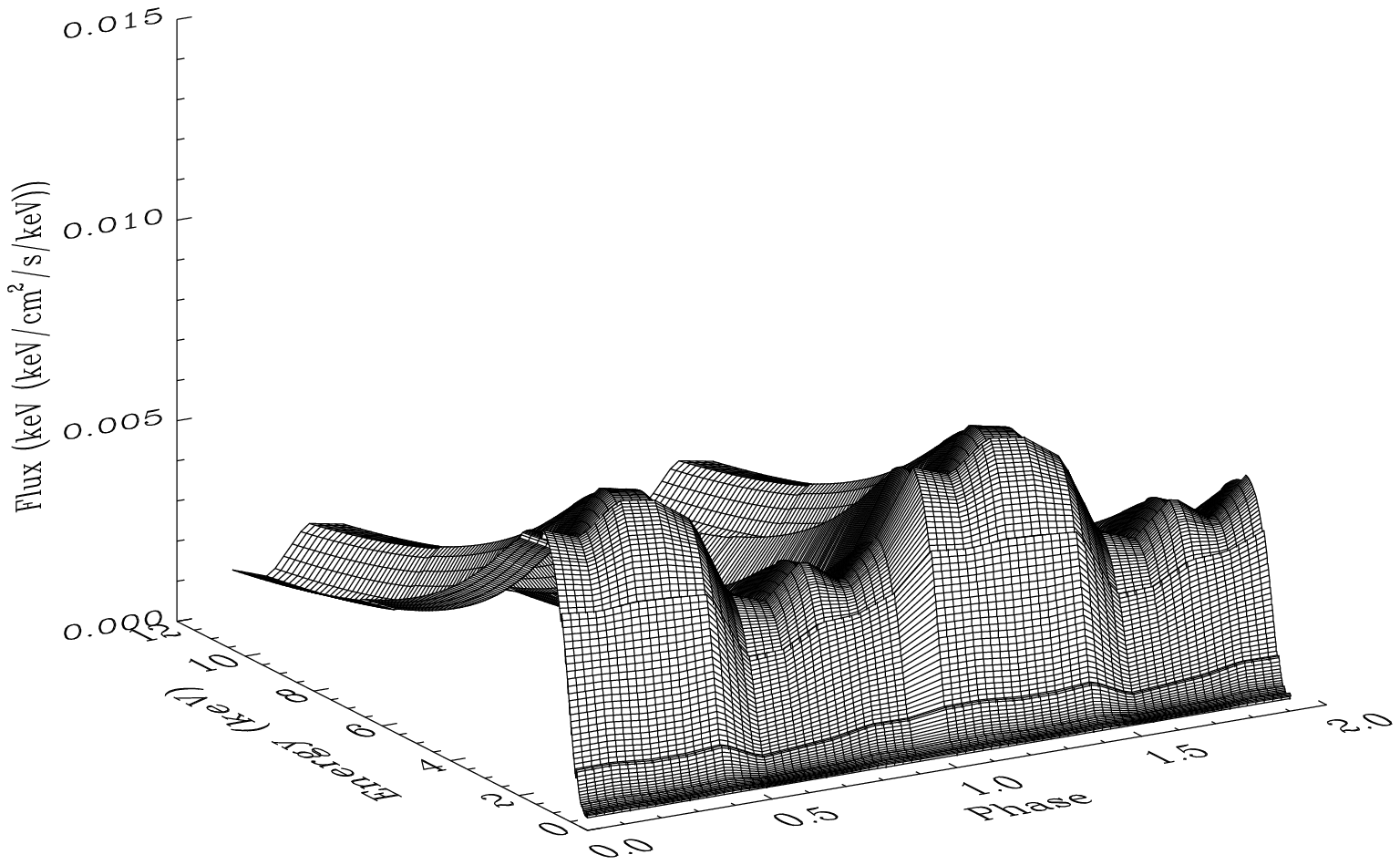,width=7cm}
\psfig{figure=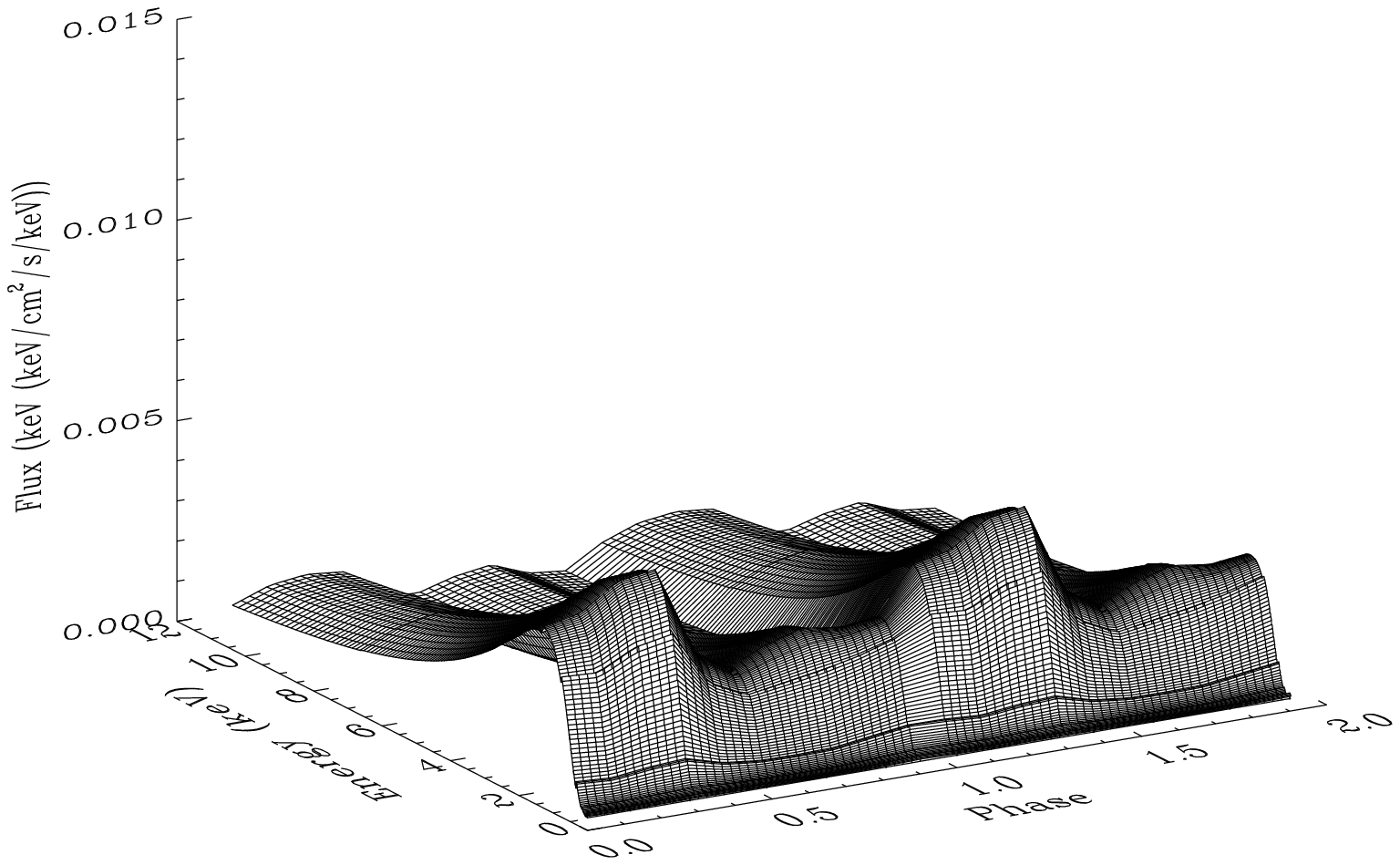,width=7cm}}}

\caption{3D pulse profiles for the five \XMM\, observations of \sgrnew
  (the epoch increases from top row to bottom row). {\em Left column}:
  pulse profiles of the BB component as a function of the energy. {\em
    Right column}: pulse profiles of the PL component as a function of
  the energy.}

\label{3Dpulsenergy}
\end{center}
\end{figure*}

%%%%%%%%%%%%%%%%%%%%%%%%%%%%%%%%%%%%%%%%%%%%%%%%%%%%%%%%%%%%%%%%%

\subsection{Swift--XRT}
\label{swift}

The \swift\, satellite (Gehrels et al. 2004) includes a wide-field
instrument, the Burst Alert Telescope (BAT; Barthelmy et al. 2005),
and two narrow-field instruments, the X-Ray Telescope (XRT; Burrows et
al. 2005) and the Ultraviolet/Optical Telescope (UVOT; Roming et
al. 2005), and discovered the bursting activity of \sgrnew\, thanks to
the large field of view of the BAT camera (Holland et al. 2008;
Barthelmy et al. 2008). We briefly report here on the \swift-XRT\,
monitoring of \sgrnew, and we refer to Palmer et al. and G{\"o}{\u
  g}{\"u}{\c s} et al. (2009, in preparation) for further details on
the \swift\, observations.

Starting a few hours after the burst activation, the \swift-XRT camera
monitored \sgrnew, collecting a few tens of observations in the
following 160\,days. The XRT instrument was operated in photon
counting (PC) mode for the first two observations, and in window
timing (WT) mode for all the following observations, which ensures
enough timing resolution (1.766\,ms) to monitor the period changes of
the source. In our analysis we ignored the first two observations in
PC mode because the were highly affected by photon pile-up.

The data were processed with standard procedures using the
\textsc{ftools} task \textsc{xrtpipeline} (version 0.12.0) and events
with grades 0--2 were selected for the WT data. For the timing and
spectral analysis, we extracted events in a region of 40$\times$40
pixels. To estimate the background, we extracted the WT events within
a similar box far from the target. The event files were used to study
the timing properties of the pulsar after correcting the photon
arrival times to the barycenter of the Solar System. For the spectral
fitting (aimed at having a reliable flux measurement over the entire
outburst) the data were grouped so as to have at least 20 counts per
energy bin. The ancillary response files were generated with
\textsc{xrtmkarf}, and they account for different extraction regions,
vignetting and point-spread function corrections. We used the latest
available spectral redistribution matrix (v011) in \textsc{caldb}. We
removed the bursts from the XRT observations taking out all the
photons corresponding to intervals where the source count rate
exceeded 5 counts\,s$^{-1}$.

\subsection{ROSAT}
\label{rosat}

The {\it R\"ontgensatellit} (ROSAT; Voges 1992; Snowden \& Schmitt
1990) Position Sensitive Proportional Counter (PSPC) serendipitously
observed the region of the sky including the position of \sgrnew\
between 1992 September 21 and 24, for an effective exposure time of
4.2\,ks. An off-axis point source, 2RXP\,J050107.7+451637, was clearly
detected in the observation, the position of which is consistent,
within uncertainties, with that of \sgrnew\ as inferred by \CXO\
(Woods et al. 2008).

The ROSAT event list and spectrum of 2RXP\,J050107.7+451637 included
about 260 background-subtracted photons accumulated from a circle of
about 1\farcm7 radius (corresponding to an encircled energy of
$\sim$90\%). The source count rate is estimated to be (6.6$\pm$ 0.5)
$\times$~10$^{-2}$ counts\,s$^{-1}$ after correction for the
point-spread function and vignetting.

%%%%%%%%%%%%%%%%%%%%%%%%%%%%%%%%%%%%%%%%%%%%%%%%%%%%%%%%%%%%%%%%%%%%%%

\begin{table*}
\begin{center}
\begin{tabular}{lccccc}
\hline
\hline
 \multicolumn{1}{l}{Parameters} & \multicolumn{5}{c}{Blackbody + Power-law} \\
\hline
& \multicolumn{1}{c}{2008-08-23} &  \multicolumn{1}{c}{2008-08-29}  &
\multicolumn{1}{c}{2008-08-31} & \multicolumn{1}{c}{2008-09-02} &
\multicolumn{1}{c}{2008-09-30} \\
\hline
kT\,(keV) & $0.70\pm0.01$ & $0.69\pm0.01$ & $0.70\pm0.01$ & $0.69\pm0.01$  &  $0.66\pm0.01$ \\
BB Radius (km) &  $1.41\pm0.05$ & $1.49\pm0.05$ & $1.42\pm0.06$ & $1.39\pm0.04$ & $1.04\pm0.06$ \\
BB flux   & $2.1\pm0.1$ & $2.3\pm0.1$ & $2.15\pm0.13$ & $1.93\pm0.07$ & $0.86\pm0.11$ \\
& &  & & & \\
$\Gamma$  & $2.75\pm0.02$  &  $2.92\pm0.04$ &  $2.90\pm0.06$ & $2.96\pm0.08$ & $3.01\pm0.04$ \\
PL flux  &  $7.7\pm0.1$ &  $5.8\pm0.1$ & $5.3\pm0.2$ & $4.9\pm0.2$ & $3.3\pm0.1$\\
& &  & & & \\
Abs. Flux  &  $4.1\pm0.1$  & $3.4\pm0.2$ & $3.14\pm0.23$ & $2.8\pm0.1$ & $1.4\pm0.1$ \\
Unab. Flux &  $9.6\pm0.1$  & $8.1\pm0.2$ & $7.5\pm0.3$ &  $7.0\pm0.3$ & $4.17\pm0.11$\\
\hline
\hline
\end{tabular}
\caption{Parameters for the spectral modelling of the phase-averaged
  spectrum of \sgrnew\, with an absorbed blackbody plus a power--law
  ($\chi_{\nu}^2$ (d.o.f.) = 1.14 (838)), for all five \XMM\,
  observations. The $N_{H}$ value is
  $(0.89\pm0.01)\times10^{22}$\,cm$^{-2}$ with solar abundances from
  Anders \& Grevesse (1989). The blackbody radius is calculated at
  infinity, and assuming a distance of 5\,kpc (note that in the error
  calculation we did not consider the uncertainty in the distance). Unless
  otherwise specified, all fluxes are unabsorbed, in the 0.5-10\,keV range,
  and in units of $10^{-11}$\,erg\,cm$^{-2}$\,s$^{-1}$. Errors
  are at the 90\% confidence level.\label{tabspec}}
\end{center}
\end{table*}

%%%%%%%%%%%%%%%%%%%%%%%%%%%%%%%%%%%%%%%%%%%%%%%%%%%%%%%%%%%%%%%%%%%%%%%%%%%%%%%%

\section{X-ray Timing Analysis}
\label{timing}

We started the timing analysis by performing a power spectrum of the
first \XMM\, observation (after having cleaned the data for the
bursts; see above), and we found a strong coherent signal at
$\sim$5.76\,s, followed by 8 significant harmonics. We then refined
our period measurement studying the phase evolution within the
observation by means of a phase-fitting technique (see Dall'Osso et
al. 2003 for details). The resulting best-fit period is
P$=5.762070(3)$\,s (1$\sigma$ confidence level; epoch 54701.0
MJD). The accuracy of 3$\mu s$ is enough to phase-connect coherently
the first two \XMM\ pointings which are about 6 days apart. The
procedure was repeated by adding, each time, a single \XMM\
pointing. The relative phases were such that the signal phase
evolution could be followed unambiguously in the 5 \XMM\,
observations, and the preliminary phase-coherent solution for these
observations had a best-fit period of P$=5.7620692(2)$\,s and
$\dot{\rm P}$ = 6.8(8) $\times$ 10$^{-12}$\,s s$^{-1}$ (MJD 54701.0
was used as reference epoch; $\chi^2\sim4$ for 3 degrees of freedom,
hereafter d.o.f.).

To better sample the pulsations in the time intervals not covered by
\XMM\ data, and increase the accuracy of our timing solution, we also
included the \su-XIS observation (Enoto et al. 2009) and part of the
\swift-XRT monitoring dataset. A quadratic term in the phase evolution
is required starting about one month after the \swift-BAT onset, when
the pulse phases increasingly deviate from the extrapolation of the
above P$-\dot{\rm P}$ solution (see Fig.\,\ref{phase}), resulting in
an unacceptable fit ($\chi^2 \sim$ 110 for 16 d.o.f.).  Therefore, we
added a higher order component to the above solution to account for
the possible presence of a temporary or secular $\ddot{\rm P}$
term. The resulting new phase-coherent solution had a best-fit for
$P=5.7620695(1)$\,s, $\dot{\rm P}$ = 6.7(1) $\times$ 10$^{-12}$\,s
s$^{-1}$, and $\ddot{\rm P}$ = -1.6(4) $\times$10$^{-19}$\,s s$^{-2}$
(MJD 54701.0 was used as reference epoch; 1$\sigma$ c.l.; $\chi^2 =
58$ for 45 d.o.f.), or $\nu$ = 0.173548754(4) Hz, $\dot{\nu}$ =
-2.01(3) $\times$ 10$^{-13}$\,Hz s$^{-1}$, and $\ddot{\nu}$ =
5(1)$\times$ 10$^{-21}$\,Hz s$^{-2}$. The time residuals with respect
to the new timing solution are reported in Fig.\,\ref{phase} (central
panel; empty squares). The significance of the inclusion of the cubic
term is 5.3$\sigma$.  Moreover, the new timing solution implies a root
mean square variability of only 0.04\,s. We note that the new timing
solution is in agreement with that reported by Israel et al. (2008a).

The negative sign of $\ddot{\rm P}$ implies that the spin-down is
decreasing on a characteristic timescale of about half a year. This
might imply that a transient increase of the spin-down above the
secular trend occurred in connection with the outburst onset, and that
the source might now be recovering toward its secular spin-down. We
note that timing components of similar strengths and with similar
evolution timescales were detected in other AXPs and SGRs following
the occurrence of glitches (Dall'Osso et al. 2003; Dib et al. 2008).
This finding suggests that a similar event might have occurred
connected to the burst and/or outburst behavior displayed by \sgrnew\,
in August 2008. Correspondingly, assuming that the secular spin-down
was an order of magnitude smaller than the one we measured during the
outburst, our findings imply a magnetic field strength of the dipolar
component in the range $7 \times 10^{13} < B_d < 2 \times
10^{14}$\,Gauss (assuming a neutron star moment of inertia of
$10^{45}$g\,cm$^{2}$).

The 0.3-11\,keV \sgrnew\ pulse profiles are relatively complex, with
several sub-peaks, though dominated by the sinusoidal fundamental
component (see Fig.\,\ref{pfenergy} and top panels of
Fig.\,\ref{pps}). The fundamental pulsed fraction calculated as
$(max-min)/(max+min)$ is fairly constant in time (although with some
oscillations) changing from 41\%$\pm$1\% during the first \XMM\
pointing, to 35\%$\pm$1\% (2$^{nd}$ pointing), to 38\%$\pm$1\%
(3$^{rd}$ and 4$^{th}$ pointings), and finally to 43\%$\pm$1\% (last
pointing; see also Tab. \ref{obslog}). At the same time both the shape
and the pulsed fraction change as a function of energy within each
pointing (see Fig.\,\ref{pulsenergy} and Fig.\,\ref{3Dpulsenergy}).

The \R\, photon arrival times were
corrected to the barycenter of the Solar System and a search for
coherent periodicities was performed in a narrow range of trial
periods (6.1--5.5\,s; we assumed a conservative value of
$|\dot{P}|$=6$\times$10$^{-10}$ s\,s$^{-1}$) centered around the 2008
August period. No significant peaks were found above the 3$\sigma$
detection threshold. The corresponding upper limit to the pulsed
fraction is about 50\%.

\section{Spectral Analysis}
\label{spectra}

For the spectral analysis we used source and background photons
extracted as described in \S\ref{obs}. The response matrices were
built using ad-hoc bad-pixel files built for each observation. We use
the {\tt XSPEC} package (version 11.3, and as a further check also the
12.1) for all fittings, and used the {\tt phabs} absorption model with
the Anders \& Grevesse~(1989) solar abundances and Balucinska-Church
\& McCammon~(1998) photoelectric cross-sections. We restricted our
spectral modeling to the EPIC-pn camera and used only the best
calibrated energy range\footnote{Note that in all our fittings there
  is a weak spurious absorption feature at $\sim2.2$\,keV, which is of
  instrumental nature, and due to the Au edge.}, namely 0.5--10\,keV.

%%%%%%%%%%%%%%%%%%%%%%%%%%%%%%%%%%%%%%%%%%%%%
\begin{center}
\begin{figure}
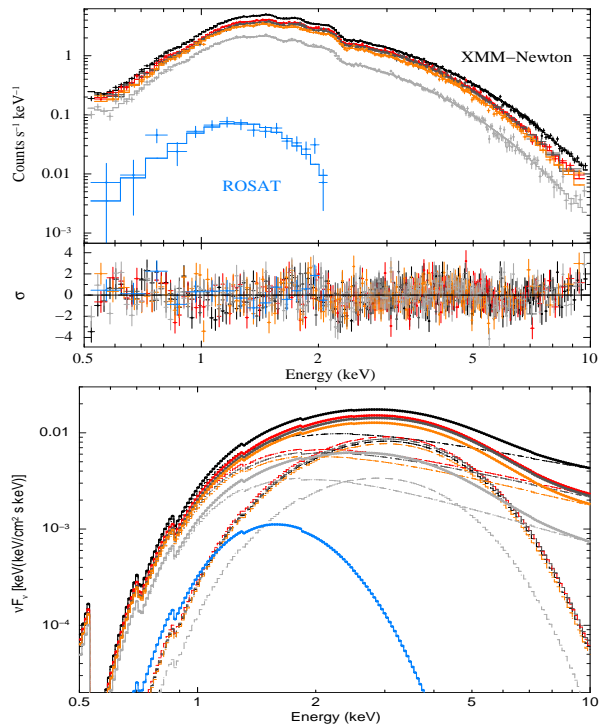

\vbox{
\psfig{figure=spectra_all_new.ps ,width=8cm,height=5cm,angle=-90}
\psfig{figure=eeufs_spectra_all.ps,width=8cm,height=4.5cm,angle=-90}}
\caption{Phase-averaged spectra and $\nu F_{\nu}$ plot of the fitted
  models for the five \XMM\, observations (again the black, red, dark
  grey, orange and light grey colors refer to the five observations
  ordered by increasing epoch) and the quiescent \R\, counterpart (in
  blue). }
\label{figspecall}
\end{figure}
\end{center}
%%%%%%%%%%%%%%%%%%%%%%%%%%%%%%%%%%%%%%%%%%%%%%%%%%%%%%%%%%%%%%%%

\subsection{Phase-averaged spectroscopy}
\label{specave}

We started the spectral analysis by fitting simultaneously the spectra
of all the \XMM\, observations with the standard blackbody (BB) plus
power-law (PL) model, leaving all parameters free to vary except for
the \nh\, which was constrained to be the same in all
observations. The values for the simultaneous modeling are reported in
Tab.\,\ref{tabspec}, with a final reduced \rchisq$=$1.14 for 838
d.o.f. (see also Fig.\,\ref{figspecall}). The values of the spectral
parameters were not significantly different when modelling each
observation separately. The measured hydrogen column density is
$N_{H}=0.89\times 10^{22}$\cm2 , and the absorbed flux in the
0.5--10\,keV band varied from 4.1 to 1.4$\times10^{-11}$\ergscm2 ,
corresponding to a luminosity range of 1.2 to 0.42$\times10^{35} \,
d_{5}^2$\, \ergs\,(where $d_{5}$ is the source distance in units of
5\,kpc; see \S~5.1 for further discussion on the source distance).

In the 0.5--10 keV band, the blackbody component accounts for
$\sim$\,15\,\% of the total absorbed flux throughout the outburst. The
blackbody radius, as derived from its normalization, is smaller than
the neutron star size, being compatible with a constant of $\sim1.4$\,km
during the first month of the outburst decay (although
hints for a decrease can be seen in the last observation). If the
blackbody emission originates from the star surface this would imply
that only a small fraction of the surface is emitting.

There is evidence that as the flux decreased, the 0.5-10\,keV spectrum
softened during the first month after the bursting activation (see
Tab.\,\ref{tabspec} and Fig.\,\ref{figspecall}). Interestingly, the BB
flux decreased much slower than the PL flux, remaining almost constant
for the first 10 days, and significantly decreasing only in the last
observation more than a month after the burst activation (see also
\S\,\ref{discussion}, Fig.\,\ref{3Dpulsenergy} and Fig.\,\ref{figspecall}).

\begin{center}
\begin{figure}
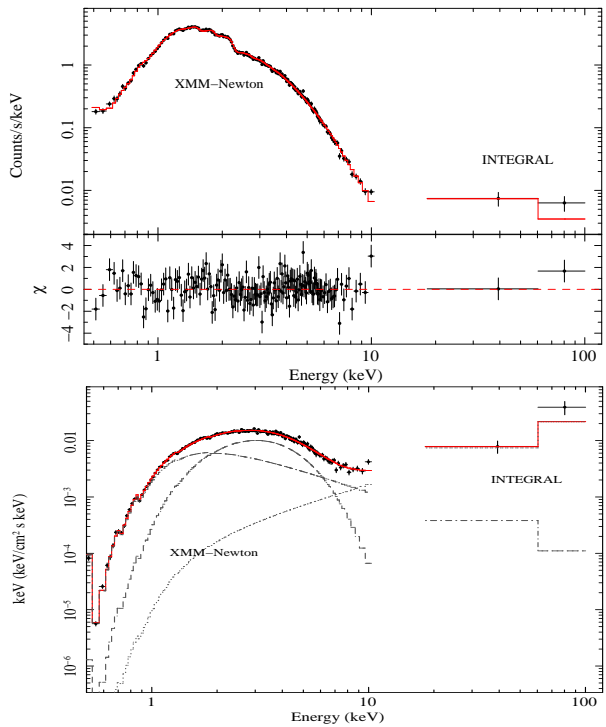

\vbox{
\psfig{figure=spec_xmm_integral_obs2_new2.ps,width=8cm,height=5cm,angle=-90}
\psfig{figure=eeufs_xmm_integral_obs2_new2.ps,width=8cm,height=4.5cm,angle=-90}}
\caption{Phase-averaged spectra of the second \XMM\, observation and
  the quasi-simultaneous \INT\, one, modelled with a blackbody plus two power-laws (see also Tab.
  \ref{tabspecxmmintegral}).}
\label{figspecxmmintegral}
\end{figure}
\end{center}

Since the \INT\, observation of \sgrnew\, was almost simultaneous to
our second \XMM\, observation, we then extended our spectral modelling
to the entire 0.5--100\,keV spectrum of the 2008 August 29
observation. We found that the BB+PL model was no longer statistically
acceptable (\rchisq=1.29 for 174 d.o.f.), and that the PL used to
model the soft X-ray spectrum could not account for the emission above
10\,keV (as it is usually the case for SGRs; G\"otz et al. 2006). We
then tried more complex models. In line with other magnetar spectra
(Kuiper et al. 2006; G\"otz et al. 2006), we added a second PL to the
data to account for the hard X-ray emission.The results are reported
in Tab.\,\ref{tabspecxmmintegral} (see also
Fig.\,\ref{figspecxmmintegral}), where we also report the F-test
probability for the addition of a further component to the fit.  We
also note that an excess in the residuals at energies larger than
8\,keV was present in the first \XMM\, observation when fit with a
BB+PL model (see Fig.\,\ref{figspecall}), probably due to the presence
of the same hard X-ray component detected by \INT, which might have
been present from the beginning of the outburst. The subsequent \INT\,
observation close to the fourth \XMM\, observation almost a week
later, did not show any hard X-ray emission. Assuming (although
unlikely) that the hard X-ray spectral index did not change during the
flux decay, we can translate our non-detection in a 3$\sigma$ flux
upper limit in the 18--60\,keV band of $< 9.7\times10^{-12}$\ergs.

To take into account the presence of this hard X-ray component we also
fit the first \XMM\, observation with a BB plus two PLs, fixing the
power-law index of the hard PL at the value inferred from the \XMM\,
plus \INT\, modelling of the second observation (namely $\Gamma =
0.8$; see the first and second columns of
Tab. \ref{tabspecxmmintegral}).  The addition of this component was
barely significant, less than in the 2008 August 29, although in the
latter case the \INT\, data were crucial in the spectral modeling.  We
similarly tried to model the third \XMM\, observation adding this PL
component but in this case the addition of this further component was
not significant. As in the case of the soft X-ray component, we found
that the hard X-ray flux decreased significantly during the outburst
decay, being undetectable by \INT\, only 10\,days after the burst
activation.

Simultaneously with the second \INT\, observation, an {\em AGILE}
observation was reported in the energy range $>$100 MeV, starting on
August 31st and ending on September 10th (Feroci et al. 2008). During
the {\em AGILE} observation the source was marginally
burst-active. The {\em AGILE}-GRID gamma-ray experiment did not
detected the source, with a reported 2$\sigma$ upper limit of
13$\times$10$^{-8}$ photon~cm$^{-2}$~s$^{-1}$.  Assuming an average
photon energy of 500\,MeV, this value corresponds to
$\sim$6$\times$10$^{-2}$ keV\,(keV~cm$^{-2}$~s$^{-1}$~keV$^{-1}$),
well below the extrapolation at this energy of the \INT\, power-law
detected during the August 29th observation (prior to the {\em AGILE}
observation), that would predict a flux at 500 MeV of $\sim$10$^{3}$
keV\,(keV~cm$^{-2}$~s$^{-1}$~keV$^{-1}$).  This indicates that as in
the AXP cases (Kuiper et al. 2006), also in this SGR the presence of a
spectral cut-off at energies between 100\,keV and 100\,MeV should be
present spectrum during outburst.

%%%%%%%%%%%%%%%%%%%%%%%%%%%%%%%%%%%%%%%%%%%%%
%%%%%%%%%%%%%%%%%%%%%%%%%%%%%%%%%%%%%%%%%%%%%%%%%%%%%%%%%%%%%%%%

%%%%%%%%%%%%%%%%%%%%%%%%%%%%%%  SPEC XMM+INTEGRAL %%%%%%%%%%%%%%%%%%%%%%%%%%%%%%%%%%%%%%%%

\begin{table}
\begin{center}
\begin{tabular}{lcc}
\hline
\hline
 \multicolumn{1}{l}{Parameters}  & \multicolumn{2}{c}{Blackbody + 2 Power-laws} \\
\hline
 & \multicolumn{1}{c}{2008-08-23} &  \multicolumn{1}{c}{2008-08-29}  \\
\hline
$N_{H}$ & $0.91\pm0.02$ & $0.93\pm0.03$ \\
& & \\
kT (keV) & $0.70\pm0.02$ & $0.69\pm0.04$   \\
BB$_1$ Radius (km) &  $1.4\pm0.1$   & $1.5\pm0.3$  \\
BB$_1$ flux   &  $2.2\pm0.1$ &  $2.7\pm0.1$  \\
& & \\
$\Gamma_{\rm soft}$  &  $2.92\pm0.07$   & $3.2\pm0.1$   \\
PL$_{\rm soft}$ flux  & $8.3\pm0.1$ & $7.1\pm0.2$ \\
$\Gamma_{\rm hard}$  &  0.8 frozen  & $0.8\pm0.2$   \\
PL$_{\rm hard}$ flux  & $3.5\pm0.1$ & $3.9\pm0.2$ \\
& & \\
Abs. flux &  $7.9\pm0.1$ & $6.8\pm0.3$    \\
Unab. flux &  $14.3\pm0.1$ & $12.6\pm0.3$     \\
& & \\
$\chi_{\nu}^2$ (d.o.f.) &  1.17 (204) & 1.18 (175)       \\
F-test prob. & $3.1\times10^{-5}$ & $4.1\times10^{-8}$  \\
\hline
\hline
\end{tabular}
\caption{Parameters of the spectral modelling of the phase-averaged
  spectra of the first two \XMM\, observations of \sgrnew\, with a
  blackbody plus two power-laws. For the second observation we used the
  quasi-simultaneous \INT\ data (see also Fig. \ref{figspecxmmintegral}
  and \S\ref{integral}). $N_{H}$ is in units of $10^{22}$\,cm$^{-2}$, and the
  blackbody radius is calculated at infinity, assuming a distance
  of 5\,kpc (uncertainties on the distance have not been included).  The
  blackbody and power-law fluxes are calculated in the 0.5--100\,keV
  band. Unless otherwise specified, fluxes are all unabsorbed and in
  units of $10^{-11}$\,erg\,cm$^{-2}$\,s$^{-1}$. Errors are at the 90\%
  confidence level.\label{tabspecxmmintegral}}
\end{center}
\end{table}

%%%%%%%%%%%%%%%%%%%%%%%%%%%%%%%%%%%%%%%%%%%%%%%%%%%%%%%%%%%%%%%%%%%%%%%%%%%%%%%%

We then studied the pre-outburst quiescent spectrum of \sgrnew\, as
observed by \R. The quiescent spectrum was well fit by either a BB or
PL single-component model (see Fig. \ref{figspecall}). The best-fit
parameters are N$_H$=6$^{+5}_{-3}\times$ 10$^{21}$ cm$^{-2}$ and
kT=0.38$^{+0.36}_{-0.15}$\,keV for the BB, and
N$_H$=8$^{+11}_{-4}\times$ 10$^{21}$ cm$^{-2}$ and $\Gamma >$0.6 for
the PL (reduced $\chi^2$=1.08 and $\chi^2$=1.13 for 17 d.o.f.,
respectively). The 0.1--2.4 keV observed flux is F$_X
\sim$1.4$\times$10$^{-12}$\,erg cm$^{-2}$ s$^{-1}$, corresponding to
an extrapolated 1--10\,keV fluxes of 1.3 and 4.2
$\times$10$^{-12}$\,erg\, cm$^{-2}$ s$^{-1}$ for the BB and PL models,
respectively. In analogy with the quiescent spectra of other
magnetars, and given the slightly better reduced $\chi^2$ we assume
that the BB spectral modeling is more correct.

No spectral features were detected in the phase-averaged \xmm\,
spectra, with 3$\sigma$ upper limits to the equivalent width of 45 and
65\,eV, for a Gaussian absorption line with $\sigma_{\rm line}$=5\,eV
(using the RGS spectra) and $\sigma_{\rm line}$=100\,eV (using the pn
spectra), respectively.

%%%%%%%%%%%%%%%%%%%%% PPS parameters %%%%%%%%%%%%%%%%%%%%%%%%%
\begin{center}
\begin{figure*}
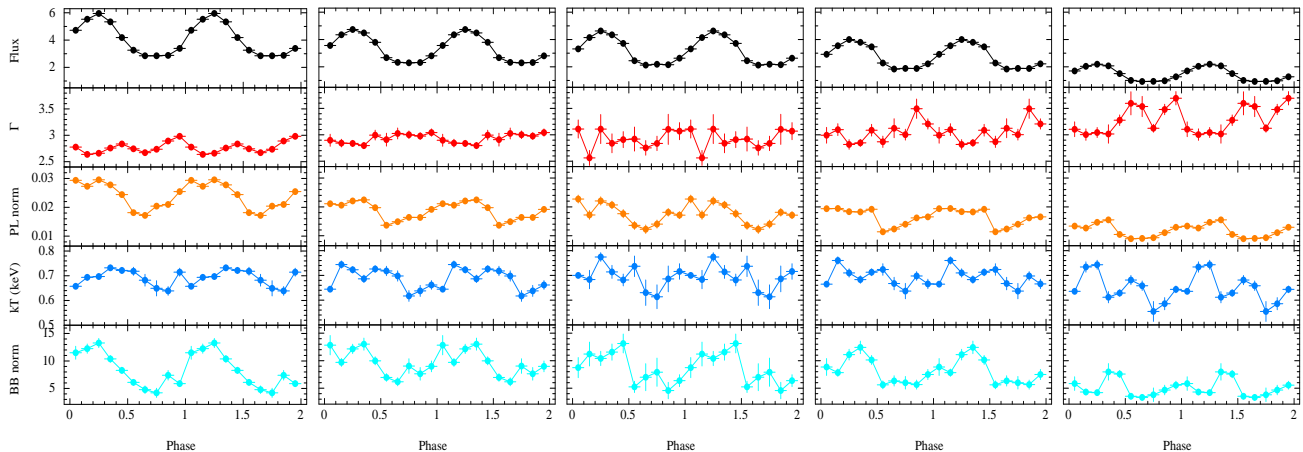


\hbox{
\psfig{figure=pps_param_bbpl_nhfix_2308new.ps,width=4.0cm,height=6.0cm,angle=-90}
\psfig{figure=pps_param_bbpl_nhfix_2908new_paoff.ps,width=3.2cm,height=5.9cm,angle=-90}
\psfig{figure=pps_param_bbpl_nhfix_3108new_paoff.ps,width=3.2cm,height=5.9cm,angle=-90}
\psfig{figure=pps_param_bbpl_nhfix_0209new_paoff.ps,width=3.2cm,height=5.9cm,angle=-90}
\psfig{figure=pps_param_bbpl_nhfix_3009new_paoff.ps,width=3.2cm,height=5.9cm,angle=-90}}

\caption{Phase-resolved spectroscopy: spectral parameters for each 0.1
  phase-bin for all five observations (epoch increases from left to
  right). For each observation all phase-resolved spectra were fitted
  simultaneously with an absorbed blackbody plus a power-law, keeping
  the $N_H$ fixed at the most accurate phase-averaged value ($N_H
  =(0.89\pm0.01)\times10^{22}$\cm2 ). Fluxes are absorbed, in the
  0.5--10keV energy range and in units of $10^{-11}$\ergscm2 .Error
  bars are at the 90\% confidence level.}
\label{pps}
\end{figure*}
\end{center}
%%%%%%%%%%%%%%%%%%%%%%%%%%%%%%%%%%%%%%%%%%%%%%%%%%%%%%%%%%%%%%%%

\subsection{Phase-resolved spectroscopy}
\label{secpps}

We performed a phase-resolved spectroscopy (PRS) for all the \xmm\,
observations. We generated 10 phase-resolved spectra for each
observation using the ephemeris reported in \S\,\ref{timing}. The
choice of the number of intervals was made {\it a priori} in order to
have enough statistics in each phase--resolved spectrum to detect, at
a 3$\sigma$ confidence level, a spectral line with an equivalent width
$>$ 30 \,eV (although none was detected). Note that given the
phase-connection of all the 5 \XMM\, observations (see
\S\ref{timing}), we can reliably follow each phase-resolved spectrum
in time.

The absorbed BB plus PL model provides excellent fits for all ten
phase-resolved spectra in all the observations, both when leaving \nh\
free and when fixing it to the most accurate value derived in the
phase-averaged fitting of all five \xmm\, observations (see
Tab.\,\ref{tabspec}). In Fig.\,\ref{pps} we have plotted the
parameters derived from the PRS analysis and compared them to the
pulse profile in each observation. All the observations showed
significant spectral variability with phase, as well as a general
softening in time. In particular, the blackbody temperature and
normalization follow the pulse profile shape rather well, and
remaining on average rather constant throughout the outburst, with a
slightly decrease in the last \XMM\, observation.  On the other hand,
the power-law parameters vary in phase and follow a more complex
behaviour, with a double-peaked change of the photon index (see also
Fig.\,\ref{3Dpulsenergy}, and \S\ref{discussion} for further discussion).

\section{Discussion}
\label{discussion}

In the last few years, thanks to the availability of wide field X-ray
instruments, as \swift-BAT, several outbursts from known AXP and SGR
have been observed, and monitored in great detail. The detection of an
outburst from \sgrnew\, has a special significance since this is the
first new SGR discovered over a decade. In this paper we presented a
comprehensive study of the spectral and timing properties of the
source in the X-rays during the entire evolution of the outburst,
starting from $\sim 1$ day after the activation and up to $\sim 160$
days later. Our investigation is based on \XMM, \swift-XRT, and \INT\
data and we also re-examined \R\ archival data in which the quiescent
emission of \sgrnew\ was detected.

\subsection{The outburst evolution and timescale}

Thanks to the \XMM\, and \swift-XRT quasi-continuous monitoring (see
\S\ref{xmm} and \S\ref{swift}), we could study in detail the flux
decay of \sgrnew\ and give an estimate of its typical
timescale. Fitting the flux evolution in the first 160\,days after the
onset of the bursting activity, we found that an exponential function
of the form Flux(t) = $K_1 + K_2\exp{-(t/t_c)}$ provides a good
representation of the data (\rchisq=1.2); the best values of the
parameters are $K_1 = (0.66\pm0.03)\times10^{-11}$\ergscm2 , $K_2
=(3.52\pm0.02)\times10^{-11}$\ergscm2 , and $t_c = 23.81\pm0.05$ days
(see Fig.\,\ref{figfluxdecay}). A fit with a power-law was not found
to be satisfactory (\rchisq=12).  Comparing the outburst decay
timescale of \sgrnew\, with other magnetars (see
Fig.\,\ref{figoutbursts}), there is a clear difference in
timescales. In particular, the outburst decays of other magnetars are
usually fitted by two components: an initial exponential or power-law
component accounting for the very fast decrease in the first day or so
(successfully observed only in a very few cases), followed by a much
flatter power-law with an index of $\delta\sim0.2-0.5$, where Flux(t)
$ = (t-t_0)^{\delta}$ (see Woods et al. 2004; Israel et al. 2007;
Esposito et al. 2008). A pure exponential flux decay with a timescale
of about 24 days is unusual and has been never observed
before. However, we caveat that the source did not reach the quiescent
level yet, hence a second component (e.g. a power-law) in the flux
decay can still appear at later times. Further monitoring observations
will allow in the future a complete modeling of the outburst decay
until the quiescent source level.

From Tab.\,\ref{tabspec} and Fig.\,\ref{3Dpulsenergy} it is apparent
that, at least in the first ten days of the outburst, the flux of the
blackbody component decayed more slowly than that of the power-law
one, both in the phase-average and the phase-resolved spectra. In
particular, fitting the phase-average BB and PL fluxes of the first 4
\XMM\, observations (see Tab.\,\ref{tabspec}) with a linear function
of the form Flux(t)=$A_1+A_2t$ we found a good fit for
$A_1(PL)=7.9(1)\times10^{-11}$\ergscm2 and $A_1(BB) =
2.2(1)\times10^{-11}$\ergscm2, and with $A_2(PL) =
-0.29(1)\times10^{-11}$\ergsscm2 and $A_2(BB) =
-0.018(3)\times10^{-11}$\ergsscm2. While the PL flux decreased by
$\sim 25\%$ from the first to the second observation (and kept
decreasing at a reduced rate in observations three and four), the BB
flux stayed approximately constant during the first four
observations. Both fluxes then substantially decreased in observation
five (see also \S\ref{secpps} and next section for the evolution of
the phase-resolved spectra). The relative decays of the thermal and
non-thermal components observed here are reminiscent of those of \wes\
after its intense burst of 2006 September 21 (Muno et al. 2007; Israel
et al. 2007). Even in that case, the PL component decayed more rapidly
than the BB flux (Israel et al. 2007).  The faster decay of the
non-thermal emission from \sgrnew\ is also corroborated by the
non-detection of the source in the second \INT\ pointing (see
\S\ref{spectra}).

%%%%%%%%%%%%%%%%%%%%%%%%%%%%%%%%%%%%%%%%%%%%%%%%%%%%%%%%%%%%%%%%%

\begin{figure*}
%\begin{center}
\hbox{
\vbox{
\psfig{figure=pps_efold_last_2308.ps,width=2.8cm,height=2.5cm,angle=-90}
\psfig{figure=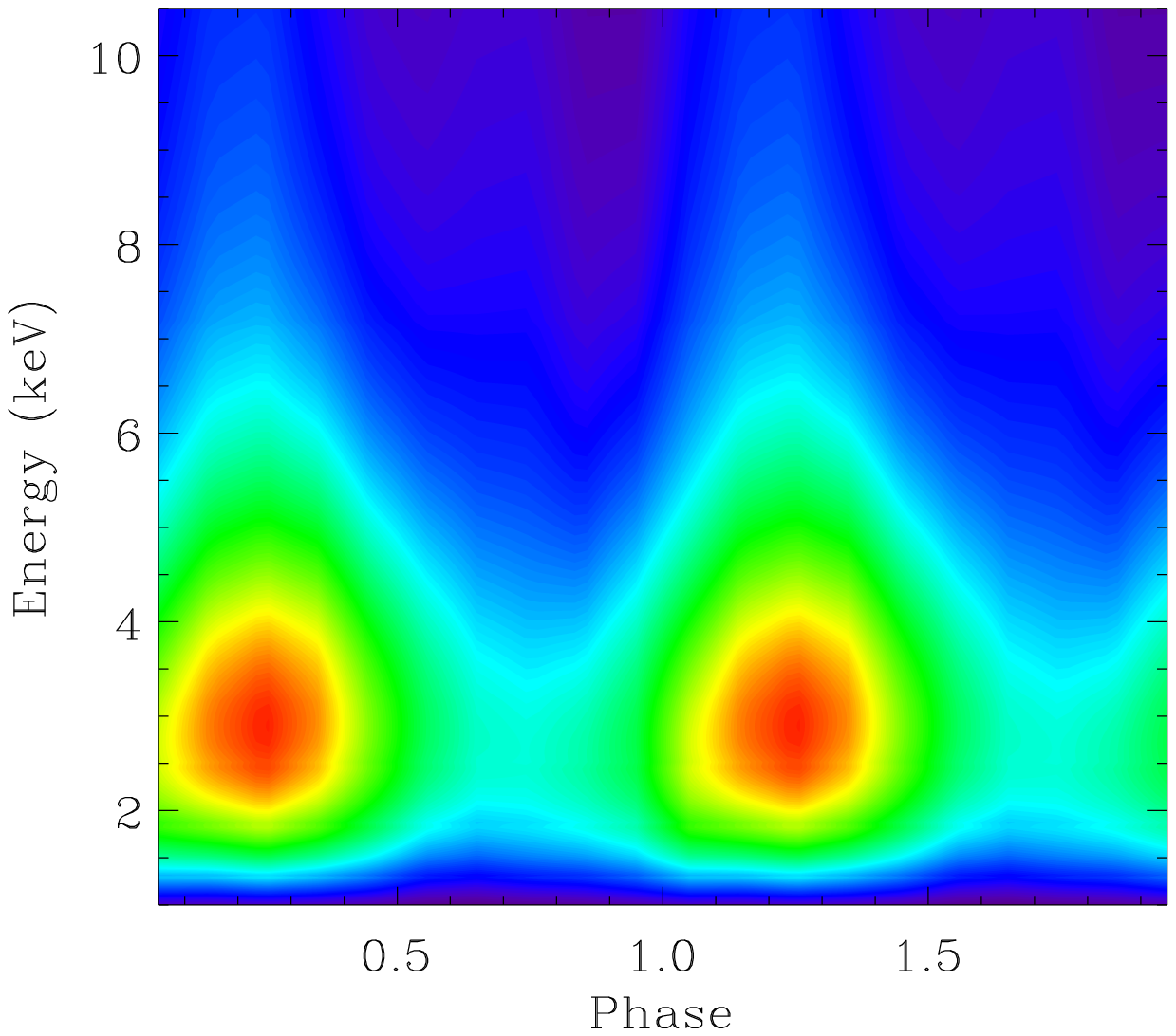,height=2.5cm,width=3.7cm}
\psfig{figure=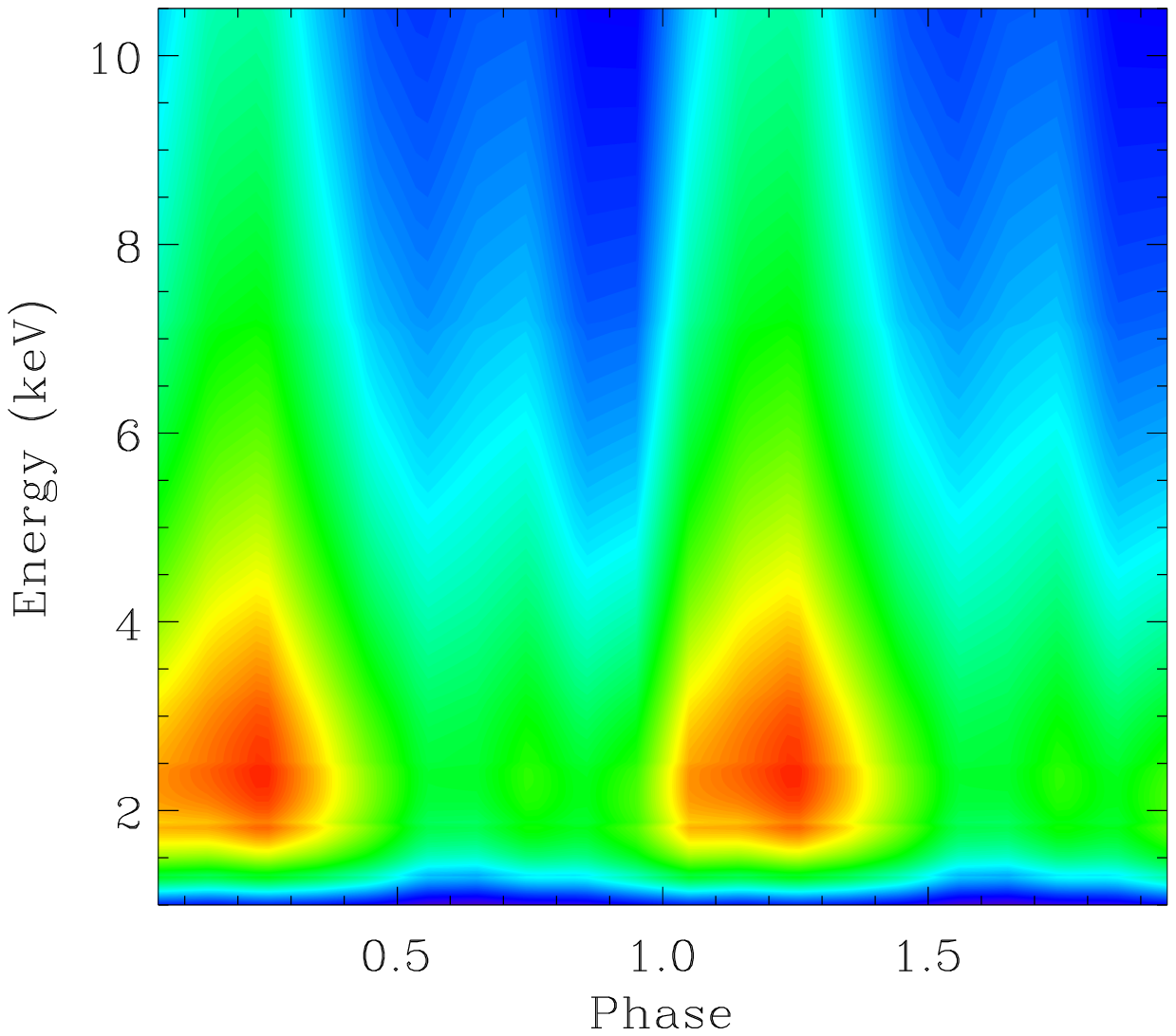,height=2.5cm,width=3.7cm}
\psfig{figure=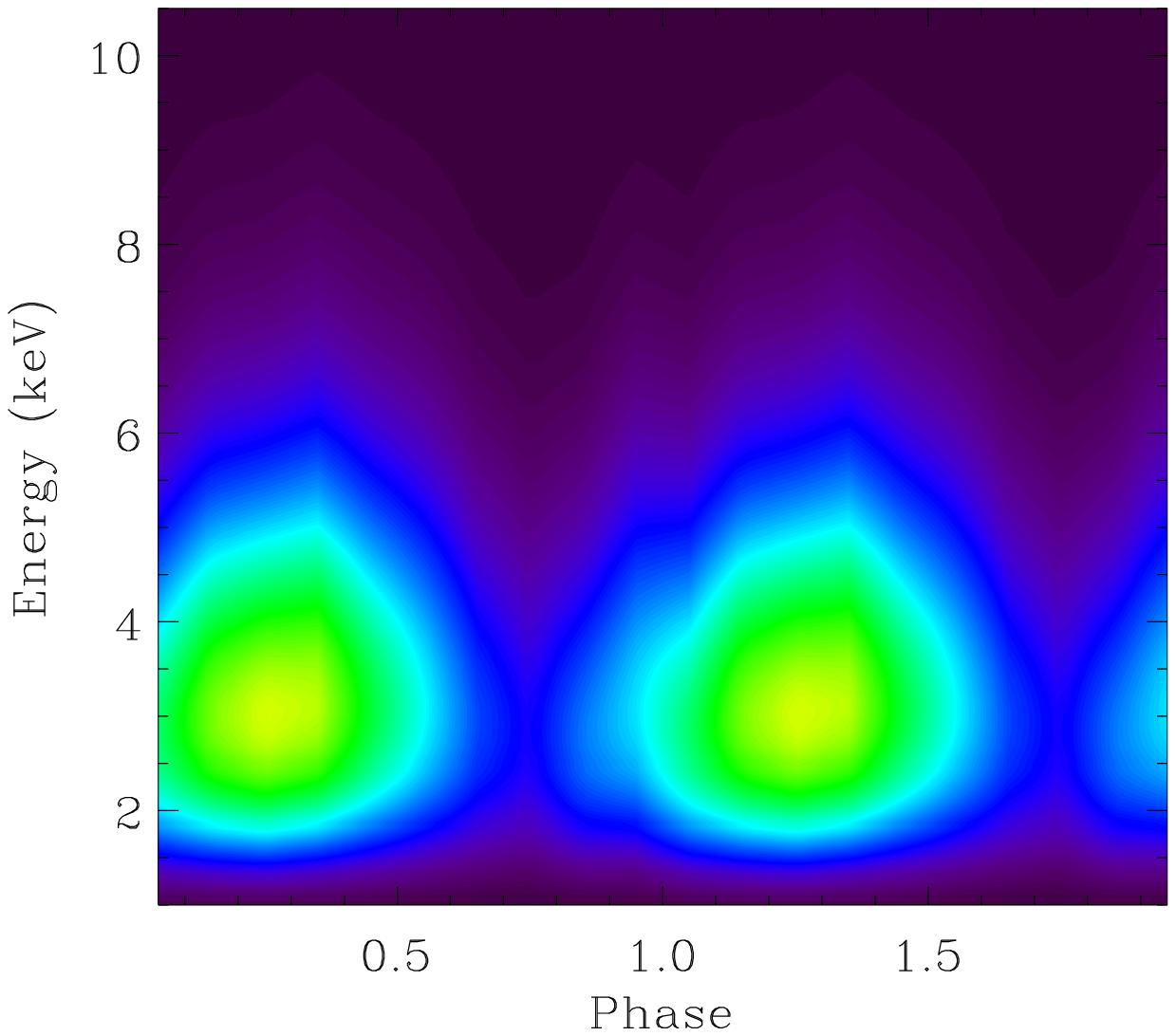,height=2.5cm,width=3.7cm}}
\vbox{
\psfig{figure=pps_efold_last_2908.ps,width=2.8cm,height=2.5cm,angle=-90}
\psfig{figure=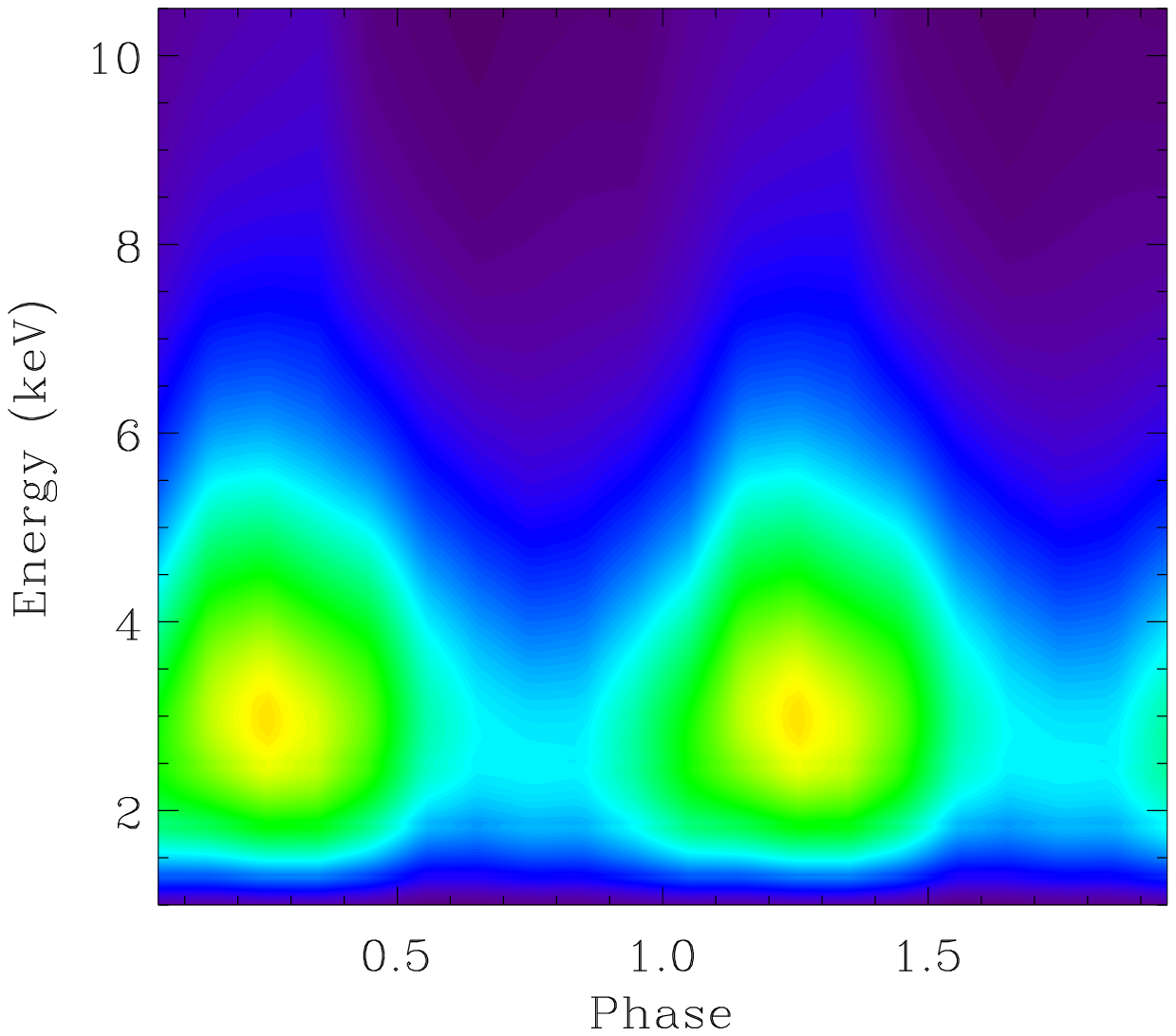,height=2.5cm,width=3.7cm}
\psfig{figure=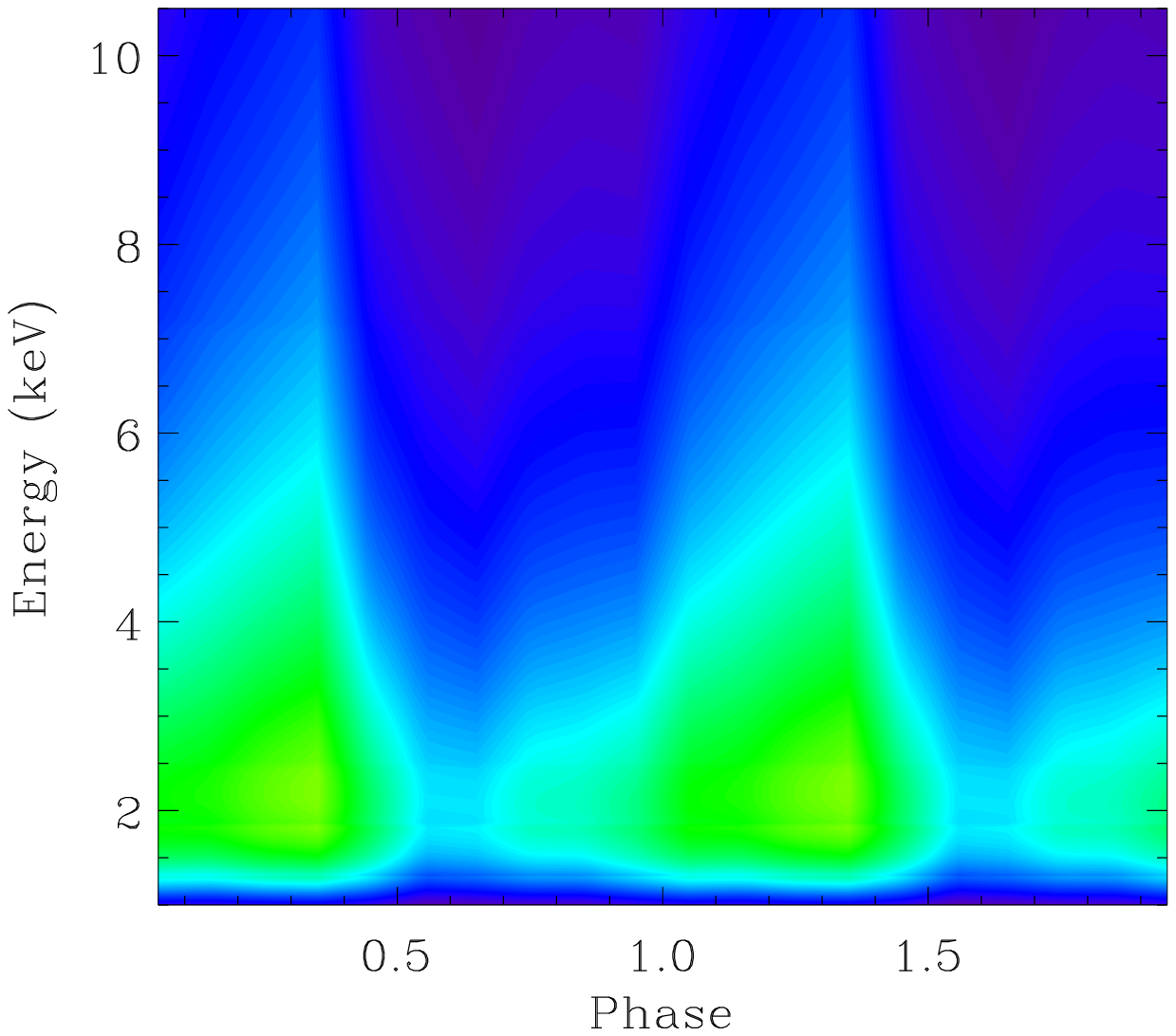,height=2.5cm,width=3.7cm}
\psfig{figure=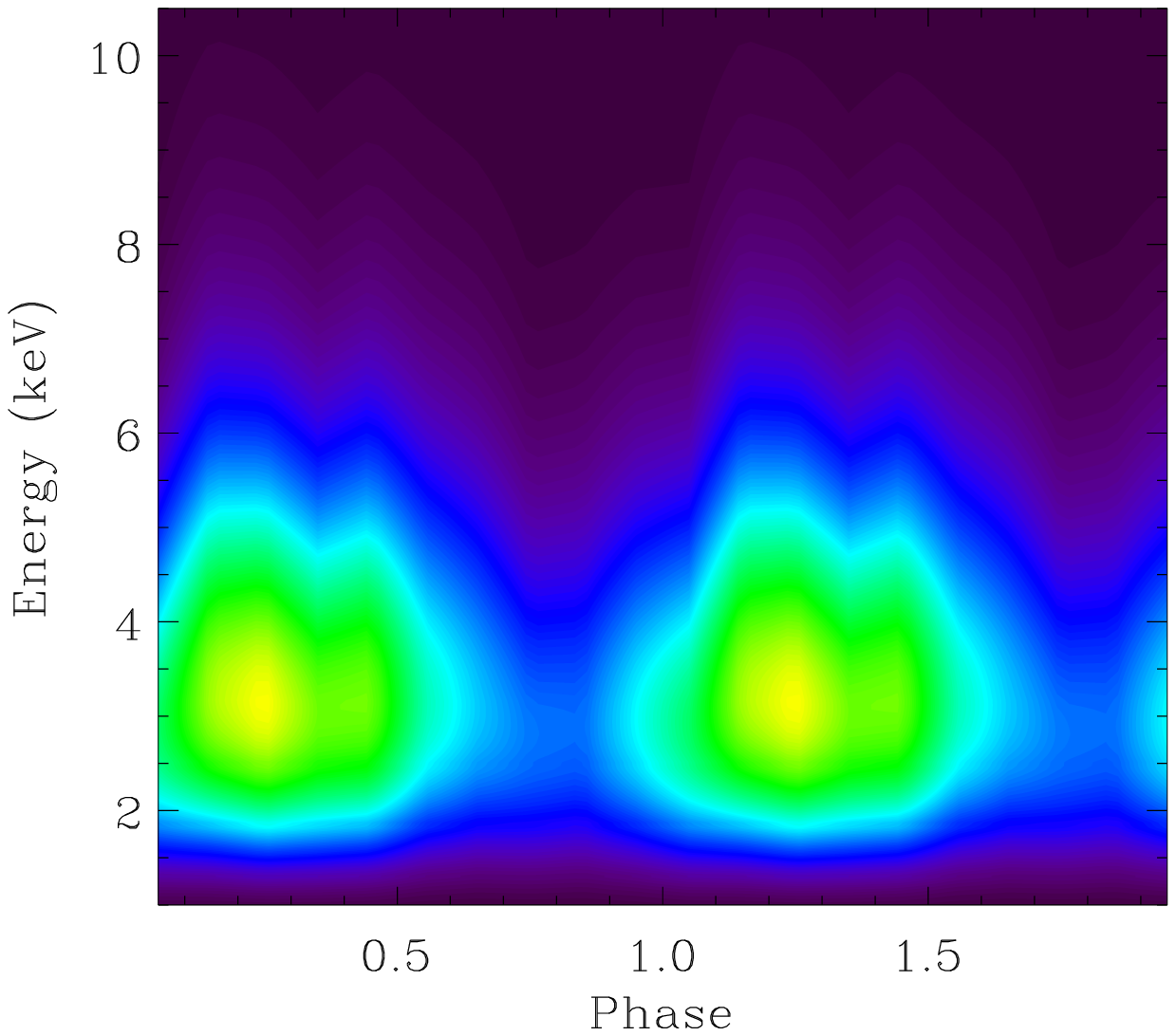,height=2.5cm,width=3.7cm}}
\vbox{
\psfig{figure=pps_efold_last_3108.ps,width=2.8cm,height=2.5cm,angle=-90}
\psfig{figure=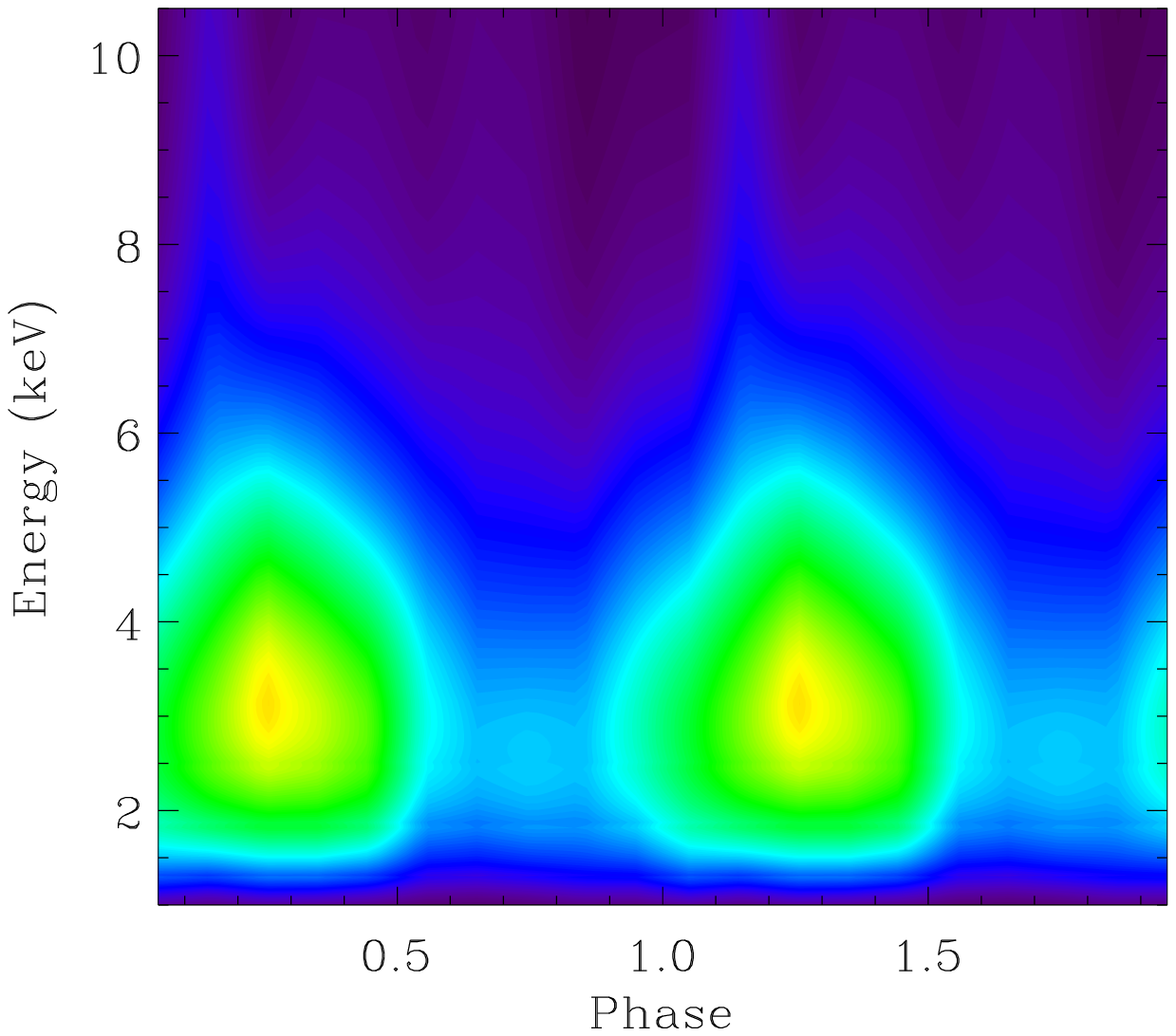,height=2.5cm,width=3.7cm}
\psfig{figure=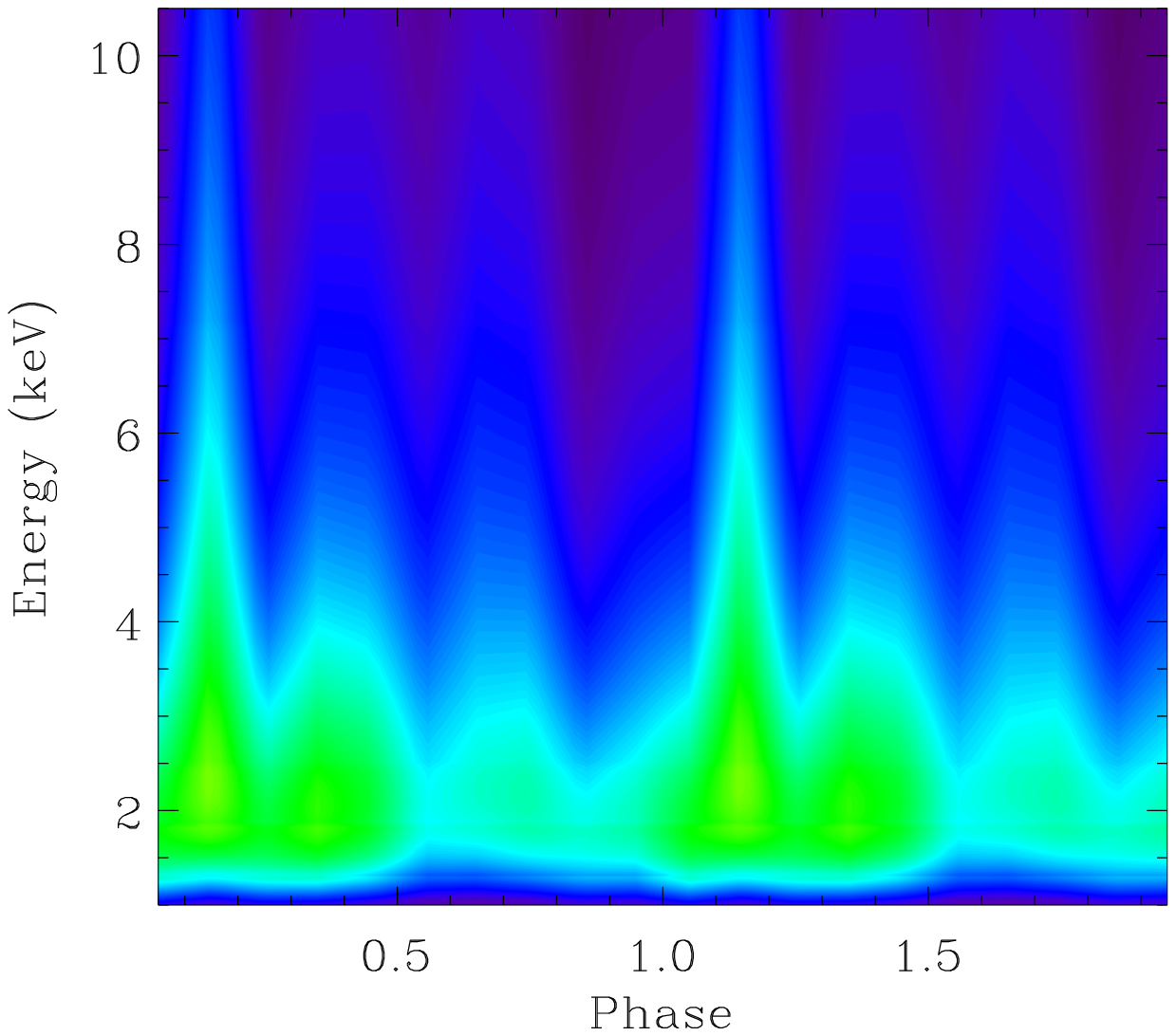,height=2.5cm,width=3.7cm}
\psfig{figure=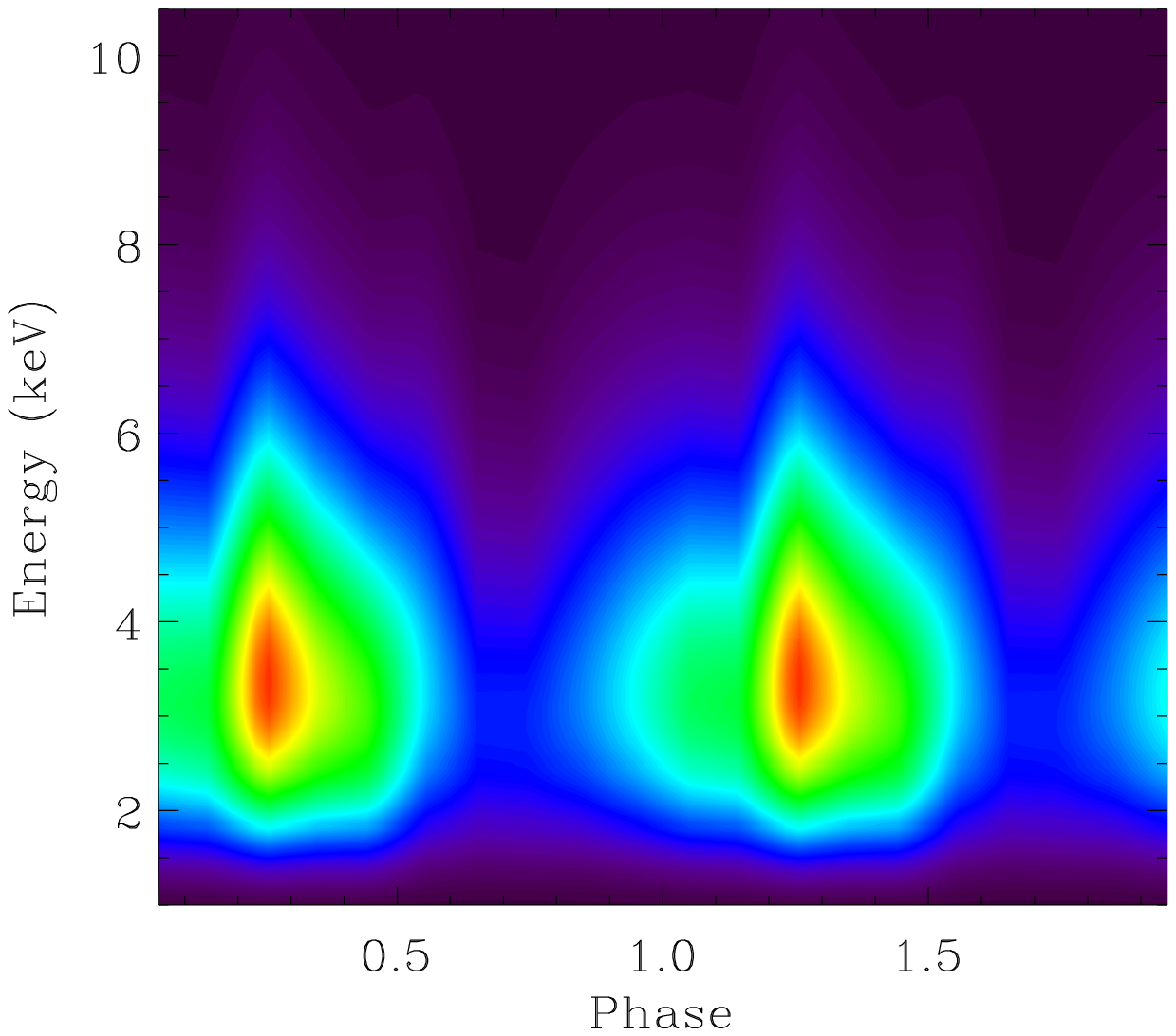,height=2.5cm,width=3.7cm}}
\vbox{
\psfig{figure=pps_efold_last_0209.ps,width=2.8cm,height=2.5cm,angle=-90}
\psfig{figure=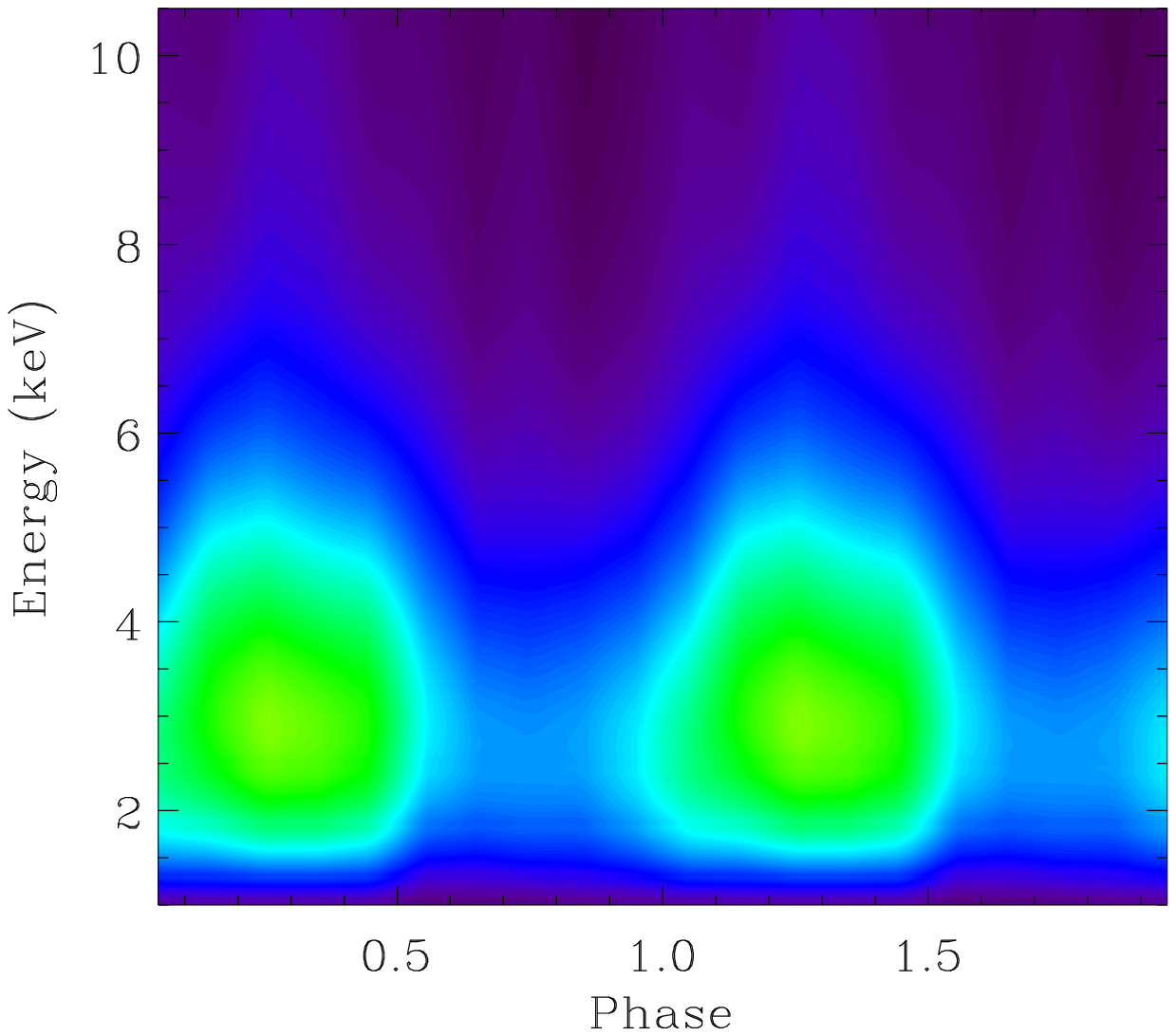,height=2.5cm,width=3.7cm}
\psfig{figure=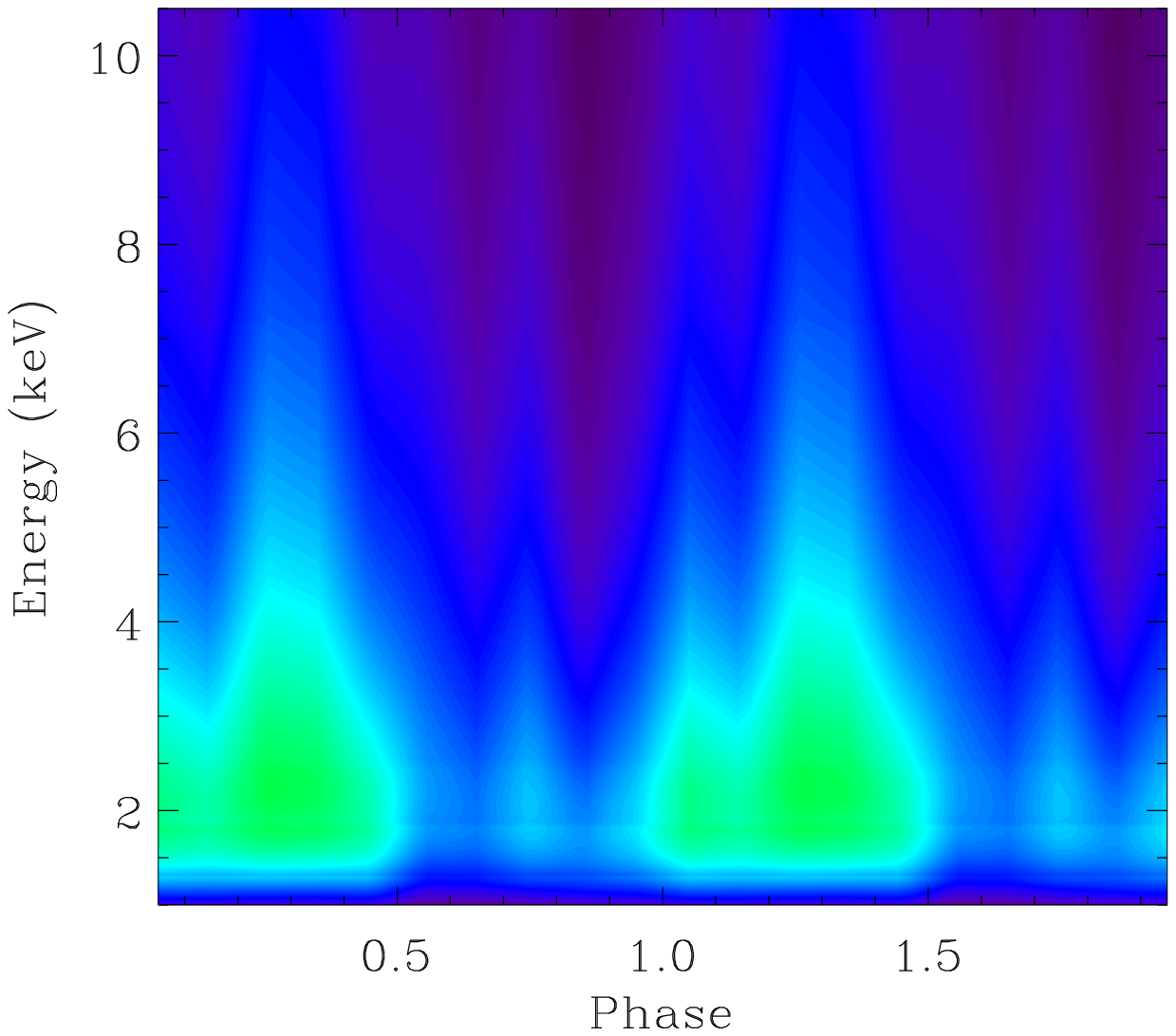,height=2.5cm,width=3.7cm}
\psfig{figure=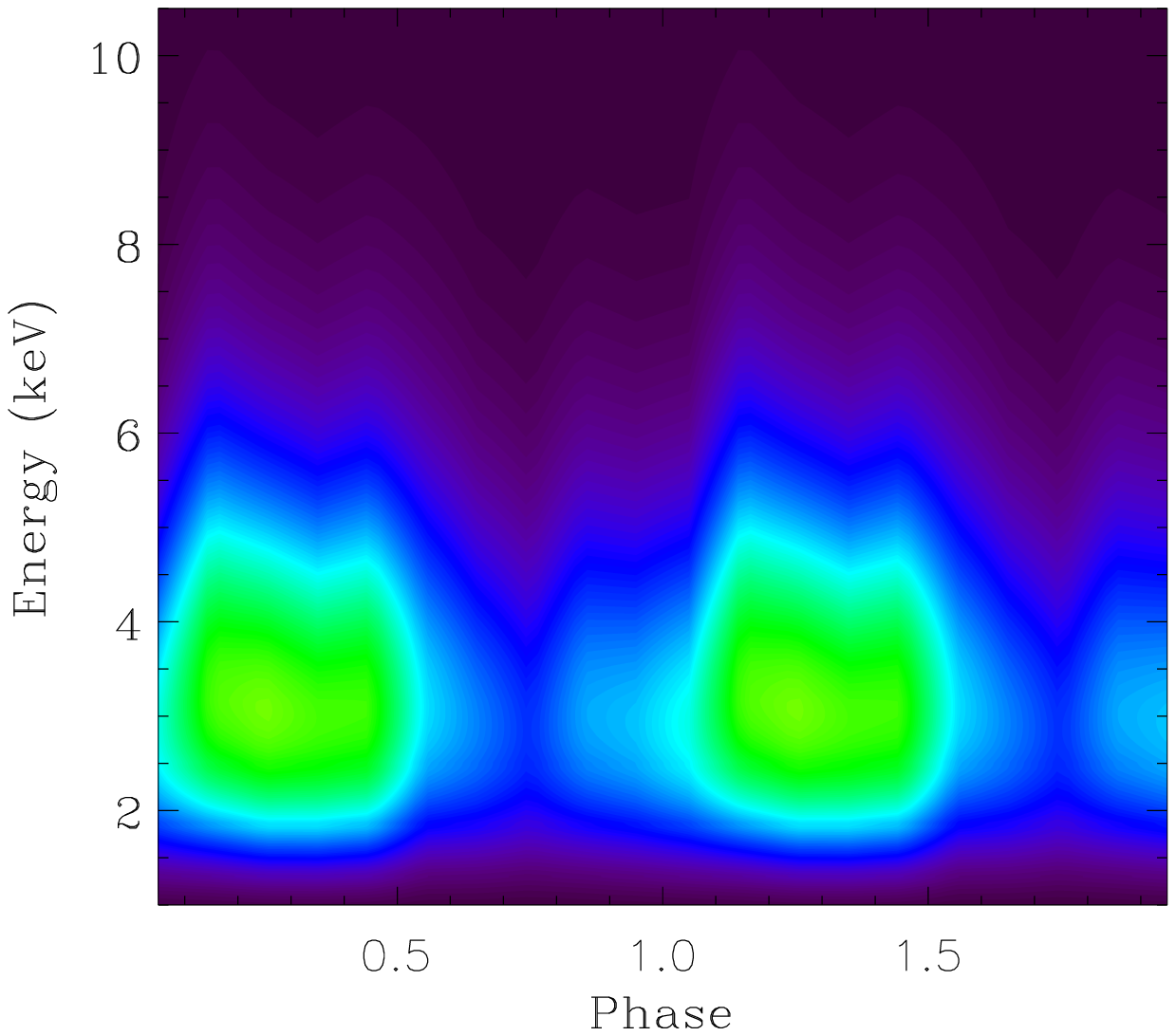,height=2.5cm,width=3.7cm}}
\vbox{
\psfig{figure=pps_efold_last_3009.ps,width=2.8cm,height=2.5cm,angle=-90}
\psfig{figure=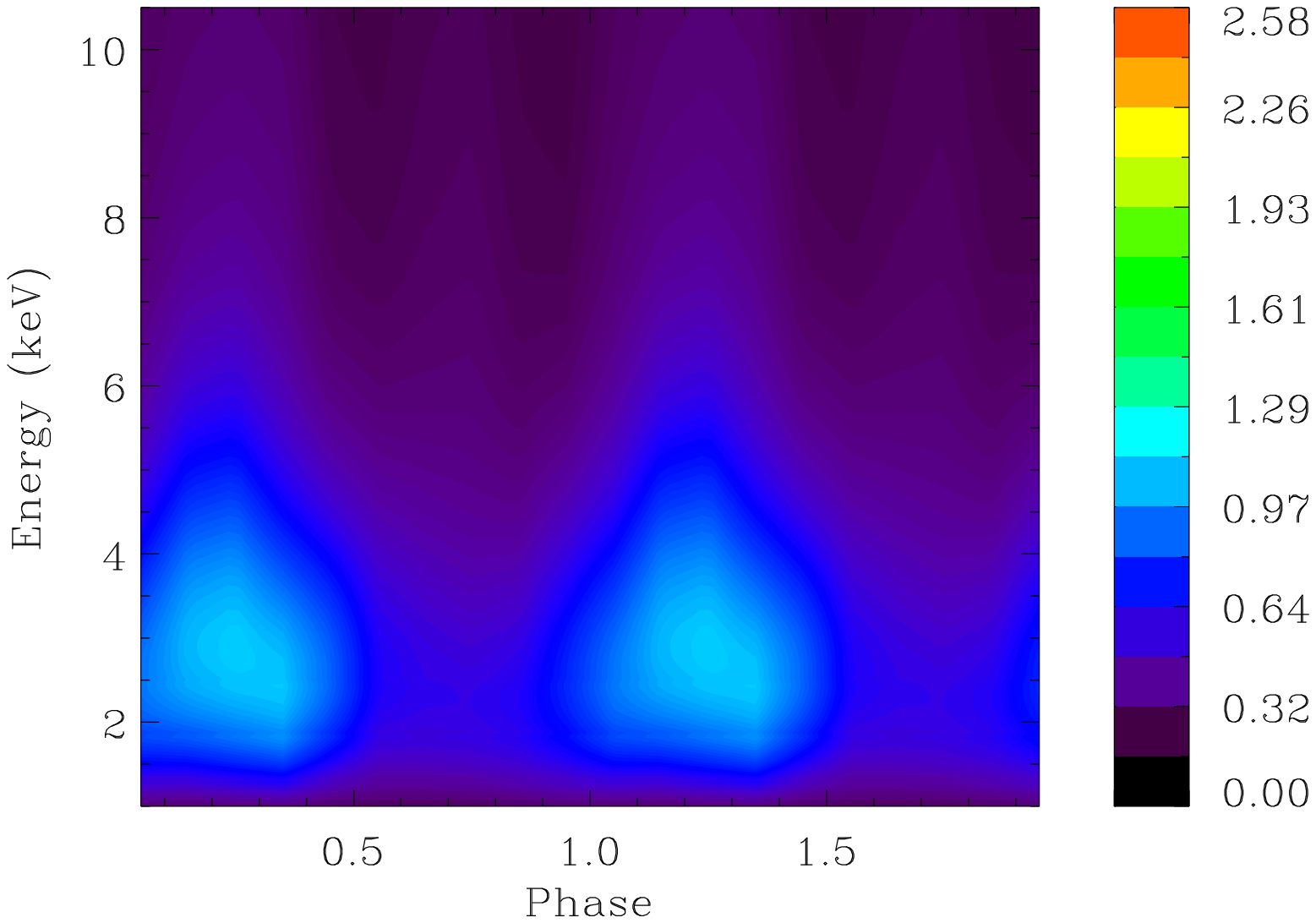,height=2.5cm,width=3.7cm}
\psfig{figure=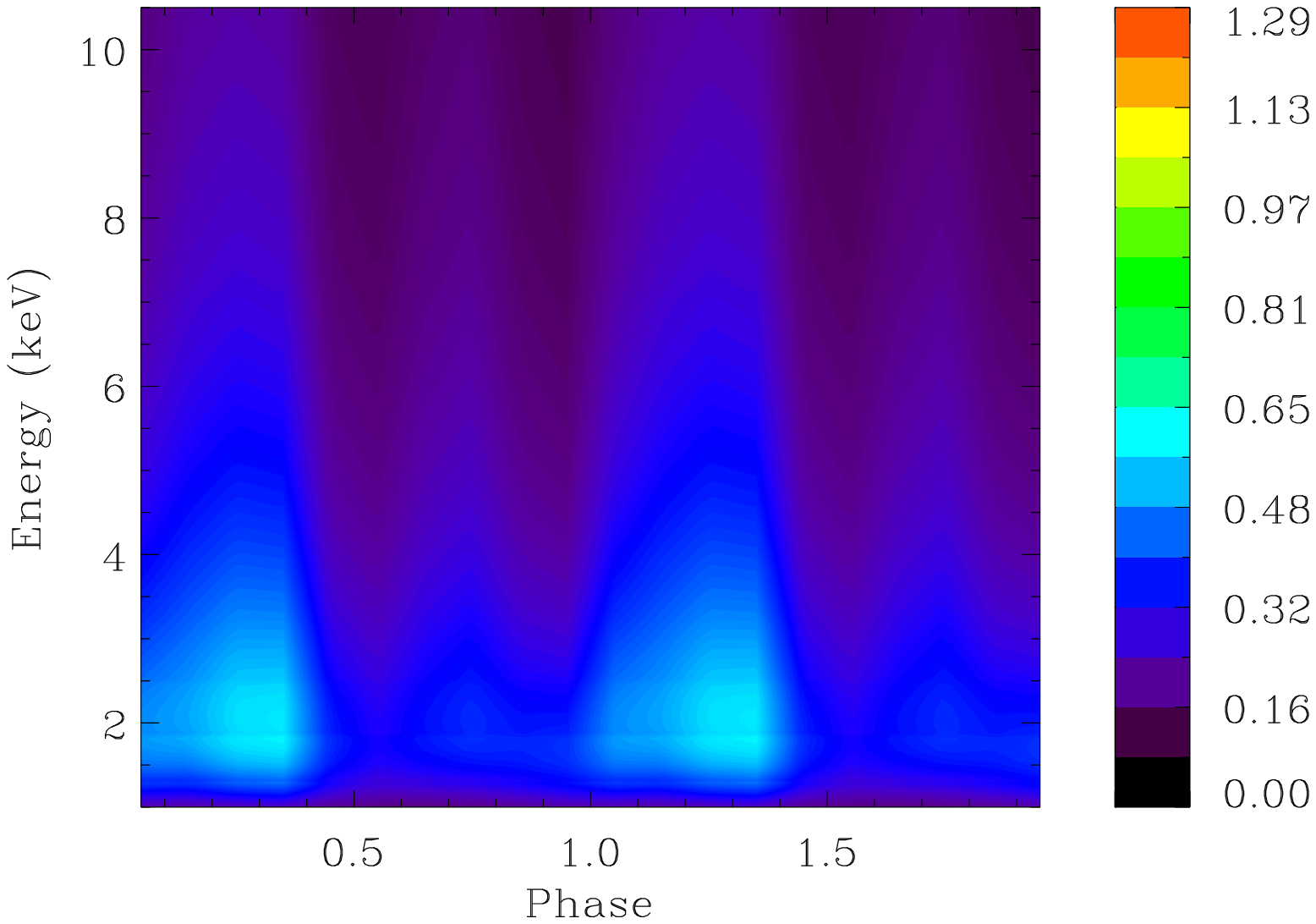,height=2.5cm,width=3.7cm}
\psfig{figure=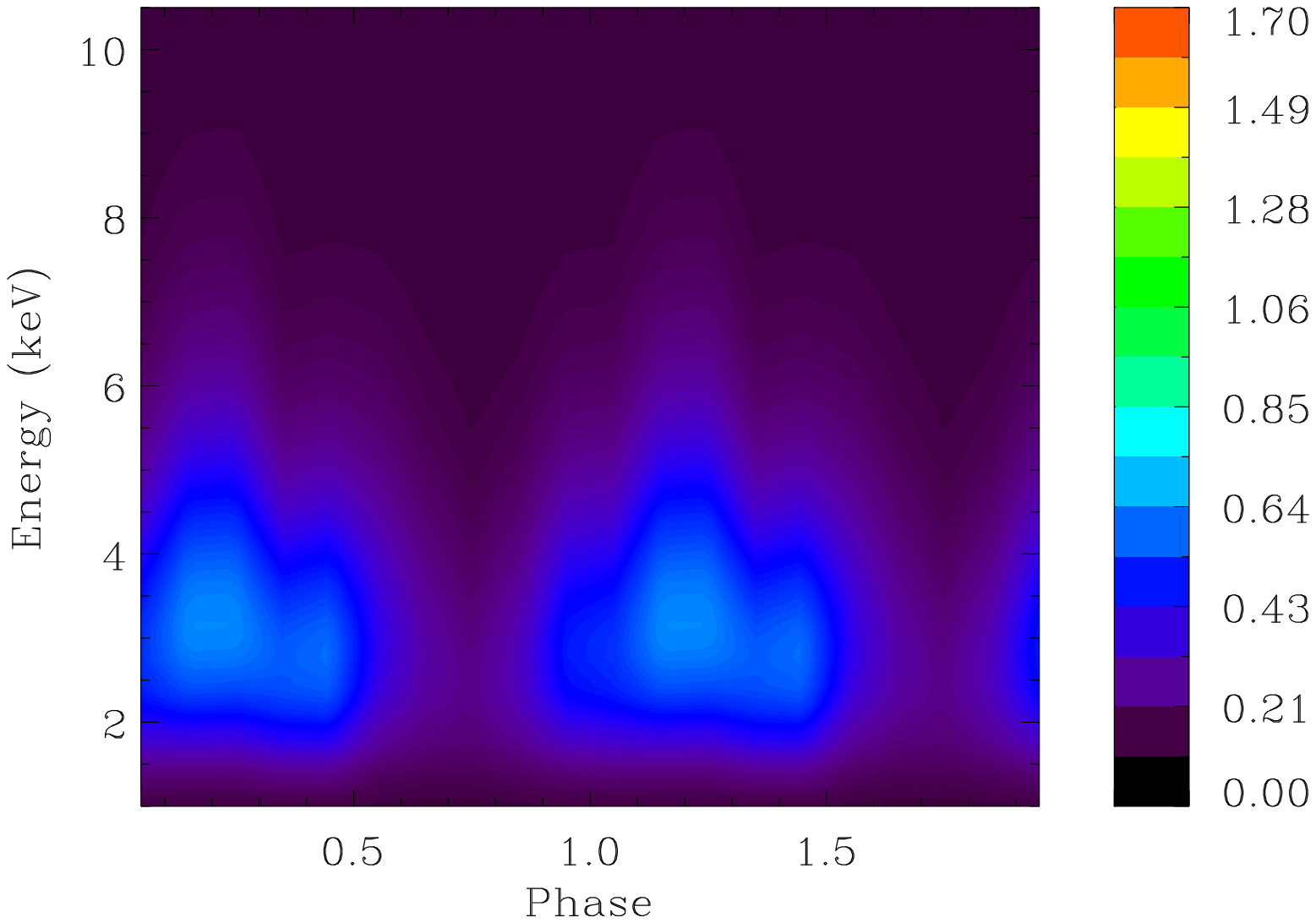,height=2.5cm,width=3.7cm}}}

\caption{Dynamic Spectral Profiles (DSPs). Each column corresponds to
  one \XMM\, observation (epoch increases from left to right: 2008
  August 23, 29, 31, September 02 and 30). For each observation, the
  top panel is the 0.3-12\,keV pulse profile, while the three bottom
  panels show in the phase/energy plane the contour plots for the total
 (second row), power-law (third row) and
  blackbody (bottom row) $\nu$F$_\nu$ flux. The colour scale is in
  units of 0.01\,keV(keV\,cm$^{-2}$\,s$^{-1}$\,keV$^{-1}$).}
\label{dsp} %\end{center}
\end{figure*}

%%%%%%%%%%%%%%%%%%%%%%%%%%%%%%%%%%%%%%%%%%%%%%%%%%%%%%%%%%%%%%%%%%%%%%

The transient character of the hard component we detected at the
beginning of \sgrnew's outburst implies that, whatever the mechanism
is, thermal bremsstrahlung in the surface layers heated by returning
currents, synchrotron emission from pairs created higher up ($\sim
100$ km) in the magnetosphere (Thompson \& Beloborodov 2005), or
resonant up-scattering of seed photons on a population of highly
relativistic electrons (Baring \& Harding 2007), it has to be
triggered by the source activity and quickly fade in a few days. All
the previous scenarios are indeed compatible with the observed
behaviour provided that a flow of highly relativistic particles is
injected into the magnetosphere during the outburst.  Note that this
is the first time that a variable hard X-ray emission is detected for
a magnetar during an outburst.  Of course, our observations did not
allow us to distinguish between a rapid spectral softening (as
expected if the particles responsible for the emission becomes less
and less energetic) and/or an overall fading of the hard component due
to a decrease in its normalization (as expected if the spatial region
occupied/heated by such particles shrinks or if their local density
decreases).

Several investigations have suggested that the observed magnetar
spectra form in the magnetosphere, where thermal photons emitted from
the neutron star's surface undergo repeated resonant scatterings
(Thompson, Lyutikov \& Kulkani 2002; Lyutikov \& Gavriil 2006;
Fernandez \& Thompson 2007; Rea et al. 2008; Nobili, Turolla \& Zane
2008a). In this scenario, the spectral shape of the non-thermal
component in the $\sim 0.1$--10 keV band (and possibly also that at
\INT\ energies; see Baring \& Harding 2007, 2008; Nobili, Turolla \&
Zane 2008b) is governed by the amount of twist which is implanted in
 the magnetosphere as a consequence of large scale crustal motions
(star-quakes). The twist must decay, due to resistive ohmnic
dissipation, in order to support its own currents (Beloborodov \&
Thompson 2007; Beloborodov 2009) and this, in turn, implies that the
high-energy component of the spectrum has to fade. If either the
initial twist is global or, as it seems more likely, it affects only a
bundle of (closed) field lines (e.g. near a magnetic pole), the
magnetosphere evolves in such a way as to confine the current-carrying
($\nabla\times\mathbf B\neq 0$) field lines closer to the magnetic
axis (Beloborodov 2009).  This necessarily quenches resonant
up-scattering because the value of the cyclotron energy in most of the
region occupied by the current-carrying field lines (which now extend
to large radii) drops below $\approx 1$ keV, the typical energy of
thermal photons.

Thompson, Lyutikov \& Kulkani (2002) and Beloborodov \& Thompson
(2007) pointed out that the surface of a magnetar with a twisted
magnetosphere is heated by the returning currents. If the twist
decays, the luminosity and the area of the heated surface decrease in
time. However, while the thermal component is expected to survive over
the timescale necessary to dissipate the twist energy, the non-thermal
component is more short-lived, since resonant scattering is no longer
possible when the current-carrying bundle becomes too small. By
comparing the theoretical expectations for a typical twist duration
and luminosity, Beloborodov (2009) found an overall agreement with the
observed properties of the transient AXP \xte, provided that the twist
was localized. In the case of \sgrnew, the typical derived evolution
time ($\sim 1$ month) requires both a twist confined to a small volume
(angular extent $\sin^2\theta\sim 0.1$) and a modest twist angle
($\psi\sim 0.1$). The distance of \sgrnew\, is not known yet, but it
has recently been estimated to be $\sim 1.5$ kpc at the lowest
(Aptekar et al. 2009), which implies a minimum source peak luminosity
$L \ga 2.5\times 10^{34}$\,\ergs.  under this case the values of the
magnetospheric parameters derived above from the timescale of the
outburst evolution are too small to explain the observed luminosity in
terms of dissipation of the twist energy alone ($L_{twist}\sim
10^{33}$\,\ergs ), and the problem worsens if the source distance is
larger (unless the emission has a beaming factor $\la 0.1$).  One
possibility is that part of the energy has been released impulsively
in the crust because of the dissipation of the toroidal field
following the star-quake, as suggested to explain the decay of \sgra\,
and \sgrd\, (Lyubarsky, Eichler \& Thompson 2003; Kouveliotou et
al. 2003). However, this scenario predicts a power-law luminosity
decline, $L\sim (t-t_0)^\delta$, which is not observed in \sgrnew. We
note that the flux decay may follow different laws in the untwisting
magnetosphere model of Beloborodov (2009), and the observed different
decay timescales of the thermal and non-thermal components fits in the
latter scenario.

\subsection{Spectral variability with phase}
\label{ppsdisc}

To study the pulse profiles and the spectral changes in phase and time
as a whole, we produced what we define hereafter as Dynamic Spectral
Profiles (DSPs), which are shown in Fig.\,\ref{dsp}. Each column in
Fig.\,\ref{dsp} is for one of the 5 \XMM\, observations (epoch
increases from left to right).  Each panel shows a contour plot of the
$\nu F_{\nu}$ flux as a function of phase and energy, and has been
derived from the 10 phase-resolved spectra extracted as explained
above. The second row refers to the total flux, as derived from the
BB+PL model, while the third and the last rows show, respectively, the
flux of the PL and BB components.  The plots illustrate well how the
source spectrum changes as phase and time, and show a clear evolution
of the phase-dependent spectrum during the outburst. At energies above
$\sim$5\,keV the PL dominates the emission at all times. From the
DSPs, and by comparing the DSPs with the pulse profiles (see
Fig.\,\ref{dsp} top panel and also Fig.\,\ref{pulsenergy}), it is also
evident that most of the sub-peaks of the pulse profiles are related
to the PL component (this is particularly evident in the third and
fourth \XMM\, observations). On the other hand, the main component of
the profiles is dominated by the BB component, which is always in
phase with the main peak. Moreover, by looking at Fig.\,\ref{dsp} it
is again evident how the PL component decreases in intensity on a
faster timescale than the BB component in all phases. Actually the BB
component is not only rather constant over the first four observations
(covering the first 10\,days after the bursting activation), but in
some phases shows a re-brightening (see Fig.\,\ref{3Dpulsenergy}, and
the third panel in the last row of Fig.\,\ref{dsp}). This is likely
due to some late heating of the surface, e.g. by returning currents.

The strong phase dependence of the non-thermal component may be
explained by the fact that, in the twisted magnetosphere model, both
the spatial distributions of the magnetospheric currents (which act as
a ``scattering medium'') and the surface emission induced by the
returning currents (which acts as source of seed photons for the
resonant scattering) are substantially anisotropic. Even under the
simple assumption where the magnetosphere is dipolar and globally
twisted, the heated part of the surface and the magnetospheric charges
cover two different ranges of magnetic colatitude. If the twist angle
varies during the outburst evolution, both distributions would move
away or toward the poles but at different rates. Of course, the
situation is more complicated if the magnetospheric twist affects a
limited bundle of field lines, as observations seem to indicate in
\sgrb\ (Woods et al. 2007) and in the transient AXP \xte\, (Perna \&
Gotthelf 2008; Bernardini et al. 2009). Recent spectral calculations
have shown the resonant comptonization in locally twisted multipolar
fields can give rise to a hard tail which is highly phase dependent
(Pavan et al. 2009). The phase-resolved spectral evolution of \sgrnew\
is very complicated, but a possible explanation for the variations of
the PL component in terms of a magnetic field which is locally
sheared, and the shear evolves in time, seems promising.

\subsection{\sgrnew: AXP or SGR?}

For about 20 years after their discovery, SGRs and AXPs were thought
to be two distinct manifestations of highly magnetic neutron stars:
the first mainly discovered and characterized by their powerful
bursting activity, and the second recognized as bright persistent soft
X-ray emitters with spectra empirically modelled by a BB+PL, and with
little or no bursting activity. Furthermore, the discovery of hard
X-ray emission (up to about 200\,keV; Kuiper et al. 2006; G\"otz et
al. 2006) from a few members of both classes, added a further
distinction, with AXPs having hard X-ray emission modelled by a second
PL component (in addition to the BB+PL describing the soft X-ray
emission) with $\Gamma_{\rm hard}\sim 0.8-1$, while the SGR emission
was the natural extrapolation at higher energies of the PL component
modelling their soft X-ray emission ($\Gamma_{\rm hard}\sim
1.5-2.0$). Over the past 6 years, the discovery of X-ray bursts from
AXPs (Kaspi et al. 2003; Woods et al. 2004), and of BB components in
the persistent spectrum of SGRs (Mereghetti et al. 2005, 2006a),
initiated a revision of this distinction between these two classes.

In this context \sgrnew\, and \1e\, can be considered the Rosetta
stone for a final unification of SGRs, AXPs and the so called
``transient AXPs (TAXPs)'', into a single class of ``magnetars
candidates''. In fact the properties of this new SGR, as well as the
characteristics of the 2009 January 22 outburst of the AXP \1e
(Gelfand \& Gaensler 2007; Halpern et al. 2008; Mereghetti et
al. 2009; Israel et al. 2009 in prep), argue for a revision of our
definition of SGRs and AXPs. In particular, \sgrnew 's 0.5-10\,keV
spectrum during outburst, is extremely soft ($\Gamma\sim$2.8-3.0)
compared to other SGRs ($\Gamma\sim$1.5-2.0). Such a soft spectrum has
been observed in the persistent emission of SGRs only during the
"quiescent" (burst-quiet) phases of \sgrd\, and \sgrc\, (Kouvelioutou
et al. 2003; Kulkarni et al. 2003; Mereghetti et
al. 2006b). Furthermore, the spectrum of the quiescent X-ray
counterpart of \sgrnew\, (see \S\ref{rosat} and \S\ref{spectra}) is
far too soft for an SGR, while resemble the pre-outburst spectrum of
the transient AXP \xte\, (Gotthelf et al. 2004).

The name \sgrnew\, came from the strong bursting activity (see
e.g. Enoto et al. 2009; Aptekar et al. 2009) which led to its
discovery. However, bursts as bright and numerous as those observed
from this source and other SGRs, have recently been observed from the
AXP \1e\, in January 2009 (Gronwall et al. 2009; Savchenko et
al. 2009; von Kienlin \& Connaughton 2009), which emitted bursts as
powerful as a typical SGR intermediate flares (Mereghetti et
al. 2009).

Another piece of evidence for the AXP-like behaviour of \sgrnew, and
the SGR-like behaviour of \1e\, is the photon index of the variable
hard X-ray component. As shown in \S\,\ref{spectra} the photon index
we measure from the \INT\, spectrum is $\Gamma\sim0.8$, which is close
to the one reported for AXPs, while the variable hard X-ray emission
during the January 2009 outburst of \1e\, has a photon index of
$\Gamma\sim1.4-1.6$ (den Hartog et al. 2009), typical of SGRs.

%%%%%%%%%%%%%%%%%%%%%%%%%%%%%%%%%%%%%%%%%%%%%
\begin{center}
\begin{figure*}
\centerline{\psfig{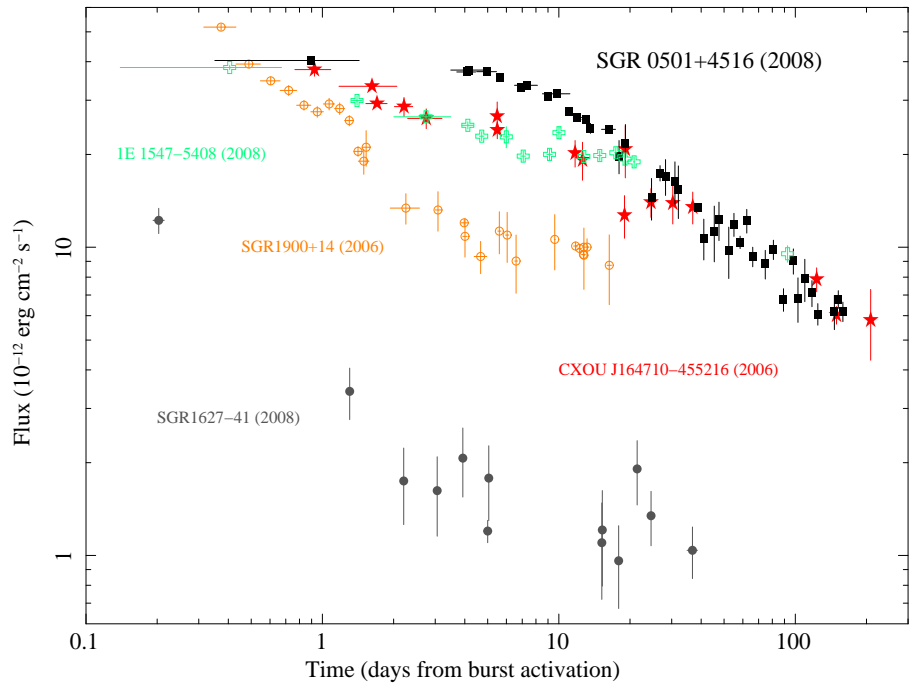}}
\caption{Flux evolution of the recent outbursts of a few magnetars
  (all observed with imaging instruments) compared with
  \sgrnew. Fluxes are the observed ones in the 1-10keV energy range,
  and the reported times are calculated in days from the detection of
  the first burst in each source. In particular we show \wes\, as red
  stars (Israel et al. 2007), \sgrd\, as grey circles (Esposito et
  al. 2008), \sgra\, as orange empty circles (Israel et al. 2008),
  \1e\, as green empty crosses (Israel et al. in prep), and \sgrnew as
  black squares (this work).}

\label{figoutbursts}
\end{figure*}
\end{center}
%%%%%%%%%%%%%%%%%%%%%%%%%%%%%%%%%%%%%%%%%%%%%%%%%%%%%%%%%%%%%%%%

\section{Summary}
\label{conclusion}

Thanks to the unprecedented prompt observational campaigns of \XMM,
\INT, and \swift, we were able to study in great detail the evolution
of the first recorded outburst from the first new SGR discovered in a
decade, \sgrnew. Furthermore, we could compare its outburst properties
with its quiescent emission as seen by \R. We found the following.

\begin{itemize}

\item Phase-connected timing analysis of the entire X-ray outburst of
  \sgrnew, strongly argue that this source is a magnetar candidate
  with a magnetic field of $B \sim 2
  \times10^{14}$\,Gauss. Furthermore, we identified a negative second
  period derivative of $\ddot{\rm P}$ = -1.6(4)$\times$10$^{-19}$\,s
  s$^{-2}$ which implies that the spin-down rate is decreasing with
  time, possibly in its way to recovering to its secular pre-outburst
  spin-down.

\item A variable hard X-ray component was detected at the beginning of
  the outburst (see Fig.\,\ref{figspecxmmintegral}), and became
  undetectable by \INT\, some time within 10\,days after the on-set of
  the bursting activity. This represent the first detection of a
  variable hard X-ray component in a magnetar over such a short
  timescale.

\item The phase-connection of all the observations allowed us to study
  the evolution in time of the phase-resolved spectra. We found that
  on top of a phase-averaged spectral softening during the outburst
  decay, with the BB component decaying on a slower timescale than the
  PL component (see Fig.\,\ref{3Dpulsenergy}), the spectral evolution also changes from phase to
  phase. The main peak of the pulse profile is dominated by the
  thermal component, while many other sub-peaks are present in the
  profiles, which are dominated instead by the non-thermal component
  (see Fig.\,\ref{dsp}).

\item No transient optical/ultraviolet source was detected by the
  Optical Monitor on board of \XMM\, (see \S\ref{obsom}).  Note that the
  optical counterpart to this source (Tanvir et al. 2008; Fatkhullin
  et al. 2008) is too faint to be observable by the OM, but we could
  constrain that no counterpart to the X-ray bursts have been observed
  with $m_{\rm UVW1}>$ 22.05.

\item From a comparison with other outbursts recently detected from
  SGRs and AXPs (see Fig.\,\ref{figoutbursts}), we show that contrary
  to other sources,  in the first 160\,days of its outburst, 
  \sgrnew\, shows a clear exponential decay on a
  rather slow timescale of about 24\,days (see Fig.\,\ref{phase}).

\item The discovery of \sgrnew, and its AXP-like characteristics,
  represents another piece of evidence in the unification of the
  magnetar candidate class, weakening further the differences between
  AXPs, TAXPs, and SGRs.

\end{itemize}

\section*{Acknowledgements}

We wish to thank Norbert Schartel for promptly approving our ToO
request for the first \XMM\, observation, the \XMM\, team for the
crucial help during the scheduling process of this monitoring program,
and the \INT\, mission operations team at ISOC and ESOC for their
support during the ToO observations. We also thank Neil Gehrels, the
\swift\, duty scientists and science planners for making the \swift\,
observations possible.  This paper is based on observations obtained
with \XMM\, and \INT , which are both ESA science missions with
instruments and contributions directly funded by ESA Member States and
the USA (through NASA), and on observations with the NASA/UK/ASI
\swift\, mission. NR is supported by an NWO Veni Fellowship and thanks
T. Enoto and K. Makishima for useful discussions on this source. PE
thanks the Osio Sotto city council for support with a G.~Petrocchi
Fellowship, SZ acknowledges STFC for support through an Advanced
Fellowship, KH is grateful to the U.S. \INT\, Guest Investigator
program for support under NASA Grant NNX08AC89G, and PU has been
supported by the Italian Space Agency through the \INT\, grant
I/008/07/0.


\begin{thebibliography}{99}

\bibitem[Anders \& Grevesse(1989)]{1989GeCoA..53..197A}
Anders, E., \& Grevesse, N. 1989, 53, 197

\bibitem[\protect\citeauthoryear{Aptekar et al.}{2009}]{apt09}
Aptekar, R.L., Cline, T.L., Frederiks, D.D., Golenetskii, S.V., Mazets, E.P., Pal'shin, V.D., 2009, ApJ, submitted (arXiv:0902.3391)

\bibitem[Balucinska et al. 1998]{ba98}
Balucinska-Church, M. \& McCammon, D. 1998, ApJ, 496, 1044

\bibitem[\protect\citeauthoryear{Baring \& Harding}{2007}]{bar07}
Baring, M.G. \&  Harding, A.K., 2007, Ap\&SS, 308, 109

\bibitem[\protect\citeauthoryear{Baring \& Harding}{2008}]{bar08}
Baring, M.G. \&  Harding, A.K., 2008, AIP Conference Proceedings, 968, 93

\bibitem[Barthelmy et al.(2005)]{2005SSRv..120..143B}
Barthelmy, S.~D., et al.\ 2005, Space Science Reviews, 120, 143

\bibitem[Barthelmy et al.(2008)]{2008ATel.1676....1B}
Barthelmy, S.~D., et al.\ 2008, The Astronomer's Telegram, 1676

\bibitem[\protect\citeauthoryear{Beloborodov \& Thompson}{2007}]{bt07}
Beloborodov A. M., Thompson C. 2007, ApJ, 657, 967

\bibitem[\protect\citeauthoryear{Beloborodov}{2009}]{bel09}
Beloborodov A. M., 2009, ApJ, in press [arXiv:0812.4873]

\bibitem[Bernardini et al.(2009)]{bern09}
Bernardini, F., et al.\ 2009, A\&A in press., arXiv:0901.2241

\bibitem[Burrows et al.(2005)]{2005SSRv..120..165B}
Burrows, D.~N., et al.\ 2005, Space Science Reviews, 120, 165

\bibitem[Chatterjee et al. (2000)]{ch00}
Chatterjee, P., Hernquist, L., \& Narayan, R. 2000, ApJ, 534, 373


\bibitem[Dall'Osso et al.(2003)]{2003ApJ...599..485D}
Dall'Osso, S., Israel, G.~L., Stella, L., Possenti, A., \& Perozzi, E. 2003, ApJ, 599, 485

\bibitem[den Hartog et al.(2009)]{2009ATel.1922....1D}
den Hartog, P.~R., Kuiper, L., \& Hermsen, W.\ 2009, The Astronomer's Telegram, 1922

\bibitem[den Herder et al.(2001)]{2001A&A...365L...7D}
den Herder, J.~W., et al. 2001, ApJ, 365, L7

\bibitem[Dib et al.(2008)]{2008ApJ...673.1044D}
Dib, R., Kaspi, V.~M., \& Gavriil, F.~P. 2008, ApJ, 673, 1044


\bibitem[\protect\citeauthoryear{Duncan \& Thompson}{1992}]{dt92}
Duncan, R., Thompson, C. 1992, ApJ, 392, L9

\bibitem[\protect\citeauthoryear{Eichler et al.}{2006}]{eichler2006}
Eichler, D., et al., 2006, arXiv:astro-ph/0611747

\bibitem[Enoto et al.(2009)]{2009ApJ...693L.122E}
Enoto, T., et al.\ 2009, ApJ, 693, L122

\bibitem[\protect\citeauthoryear{Esposito et al.}{2008}]{e08}
Esposito, P., et al., 2008, MNRAS, 390, L34


\bibitem[Fatkhullin et al.(2008)]{2008GCN..8160....1F}
Fatkhullin, T., et al.\ 2008, GRB Coordinates Network, 8160

\bibitem[\protect\citeauthoryear{Fernandez \& Thompson}{2007}]{ft07}
Fernandez R., \& Thompson C., 2007, ApJ, 660, 615

\bibitem[Feroci et al.(2008)]{2008ATel.1705....1F}
Feroci, M., et al.\ 2008, The Astronomer's Telegram, 1705

\bibitem[Gelfand
\& Gaensler(2007)]{2007ApJ...667.1111G}
Gelfand, J.~D., \& Gaensler, B.~M.\ 2007, ApJ, 667, 1111

\bibitem[Gelfand et al.(2008)]{2008GCN..8168....1G}
Gelfand, J.~D., Taylor, G., Kouveliotou, C., Gaensler, B.,  \& van der Horst, A.~J.\ 2008, GRB Coordinates Network, 8168

\bibitem[Gehrels \& Swift Team(2004)]{2004AIPC..727..637G}
Gehrels, N., \& Swift Team 2004, Gamma-Ray Bursts: 30 Years of Discovery, 727, 637

\bibitem[Gogus et al.(2008)]{2008ATel.1677....1G}
Gogus, E., Woods, P., \& Kouveliotou, C.\ 2008, The Astronomer's Telegram, 1677

\bibitem[Gotthelf et al.(2004)]{2004ApJ...605..368G}
Gotthelf, E.~V., Halpern, J.~P., Buxton, M., \& Bailyn, C.\ 2004, ApJ, 605, 368

\bibitem[\protect\citeauthoryear{G\"otz et al.}{2006}]{g06}
G\"otz, D., et al., 2006, A\&A, 449, L31


\bibitem[\protect\citeauthoryear{Gronwall et al.}{2009}]{gro09}
Gronwall, C., et al. 2009, GRB Coordinates Network, 8833

\bibitem[Halpern et al.(2008)]{h08}
Halpern, J.P., Gotthelf, E.V., Reynolds, J., Ransom, S.M., Camilo, F. 2008, ApJ, 676, 1178

\bibitem[Hessels et al.(2008)]{2008GCN..8134....1H}
Hessels, J., Rea, N., Ransom, S., \& Stappers, B.\ 2008, GRB Coordinates Network, 8134

\bibitem[Holland et al.(2008)]{2008GCN..8112....1H}
Holland, S.~T., et al.\ 2008, GRB Coordinates Network, 8112

\bibitem[\protect\citeauthoryear{Hurley et al.}{1999}]{h99}
Hurley, K., et al., 1999, Nature,  397, 41

\bibitem[\protect\citeauthoryear{Hurley et al.}{2005}]{h05}
Hurley, K., et al., 2005, Nature,  434, 1098


\bibitem[\protect\citeauthoryear{Israel et al.}{2007}]{giallo07}
Israel, G.L., et al., 2007, ApJ, 664, 448

\bibitem[Israel et al.(2008)]{2008ATel.1837....1I}
Israel, G.~L., et al.\ 2008a, The Astronomer's Telegram, 1837

\bibitem[Jansen et al.(2001)]{2001A&A...365L...1J}
Jansen, F., et al.\ 2001, ApJ, 365, L1

\bibitem[Kaspi et al.(2003)]{2003ApJ...588L..93K}
Kaspi, V.~M., Gavriil, F.~P., Woods, P.~M., Jensen, J.~B., Roberts, M.~S.~E., \& Chakrabarty, D.\ 2003, ApJ, 588, L93

\bibitem[\protect\citeauthoryear{Kaspi}{2007}]{k07}
Kaspi, V., 2007, Ap\&SS, 308, 1

\bibitem[\protect\citeauthoryear{Kouveliotou et al.}{2003}]{kou03}
Kouveliotou, C. et al.  2003, ApJ, 596, L79

\bibitem[\protect\citeauthoryear{Kuiper, Hermsen \& Mendez}{2004}]{khm04}
Kuiper, L., Hermsen, W., \& Mendez, M., 2004, ApJ, 613, 1173

\bibitem[\protect\citeauthoryear{Kuiper et al.}{2006}]{k06}
Kuiper, L., et al., 2006, ApJ, 645, 556

\bibitem[Kulkarni et al.(2003)]{2003ApJ...585..948K}
Kulkarni, S.~R., Kaplan, D.~L., Marshall, H.~L., Frail, D.~A., Murakami, T., \& Yonetoku, D.\ 2003, ApJ, 585, 948

\bibitem[Kulkarni \& Frail(2008)]{2008GCN..8130....1K}
Kulkarni, S.~R., \& Frail, D.~A.\ 2008, GRB Coordinates Network, 8130

\bibitem[\protect\citeauthoryear{Lebrun et al.}{2003}]{isgri}
Lebrun, F., Leray, J.P., Lavocat, P., et al. 2003, A\&A, 411, L141

\bibitem[\protect\citeauthoryear{Lyubarsky, Eichler \& Thompson}{2003}]{let03}
Lyubarsky, Y., Eichler, D. \& Thompson, C. 2003, ApJ, 580, L69

\bibitem[Lyutikov(2003)]{2003MNRAS.346..540L}
Lyutikov, M. 2003, MNRAS, 346, 540

\bibitem[\protect\citeauthoryear{Lyutikov \& Gavriil}{2006}]{lg06}
Lyutikov M.,  \& Gavriil F.P. 2006, MNRAS, 368, 690

\bibitem[Mason et al.(2001)]{2001A&A...365L..36M}
Mason, K.~O., et al.\ 2001, A\&A, 365, L36

\bibitem[\protect\citeauthoryear{Mazets et al.}{1979}]{m79}
Mazets, E.P., et al.,  1979, Nature,  282, 587

\bibitem[\protect\citeauthoryear{Mereghetti et al.}{2005}]{m05}
Mereghetti, S., et al., 2005, A\&A, 433, L9

\bibitem[Mereghetti et al.(2006)]{2006A&A...450..759M}
Mereghetti, S., et al.\ 2006a, ApJ, 450, 759

\bibitem[Mereghetti et al.(2006)]{2006A&A...450..759M}
Mereghetti, S., et al.\ 2006b, ApJ, 653, 1423

\bibitem[\protect\citeauthoryear{Mereghetti}{2008}]{m08}
Mereghetti, S., 2008, A\&A Review, 15, 225

\bibitem[\protect\citeauthoryear{Mereghetti}{2008}]{m08}
Mereghetti, S., et al. 2009, ApJ submitted

\bibitem[Monet et al.(2003)]{2003AJ....125..984M}
Monet, D.~G., et al.\ 2003, AJ, 125, 984

\bibitem[Muno et al.(2007)]{2007MNRAS.378L..44M}
Muno, M.~P., Gaensler, B.~M., Clark, J.~S., de Grijs, R., Pooley, D., Stevens, I.~R.,
\& Portegies Zwart, S.~F.\ 2007, MNRAS, 378, L44

\bibitem[\protect\citeauthoryear{Nobili, Turolla \& Zane}{2008a}]{ntz08}
Nobili L.,  Turolla R., \& Zane S. 2008a, MNRAS, 386, 1527

\bibitem[\protect\citeauthoryear{Nobili, Turolla \& Zane}{2008b}]{ntz2}
Nobili L.,  Turolla R., \& Zane S. 2008b, MNRAS, 389, 989

\bibitem[\protect\citeauthoryear{}{Ouyed, Leahy \& Niebergal 2007a}]{o07a}
Ouyed, R., Leahy, D., \& Niebergal, B. 2007a, A\&A, 473, 357


\bibitem[\protect\citeauthoryear{}{Ouyed, Leahy \& Niebergal 2007b}]{o07b}
Ouyed, R., Leahy, D., \& Niebergal, B. 2007b, A\&A, 475, 73

\bibitem[\protect\citeauthoryear{Palmer et al.}{2005}]{p05}
Palmer, D.M., et al., 2005, Nature, 434, 1107

\bibitem[\protect\citeauthoryear{Pavan et al.}{2009}]{lucia09}
Pavan, L., et al., 2009, MNRAS in press, arXiv:0902.0720

\bibitem[Perna et al.(2000)]{pe00}
Perna, R., Hernquist, L., \& Narayan, R. 2000, ApJ, 541, 344

\bibitem[\protect\citeauthoryear{Rea et al.}{2008}]{rea08}
Rea, N., Zane, S., Turolla, R., Lyutikov, M., G\"otz, D. 2008, ApJ, 686, 1245

\bibitem[Rea et al.(2008b)]{2008GCN..8159....1R}
Rea, N., Rol, E., Curran, P.~A., Skillen, I., Russell, D.~M., \& Israel, G.~L.\ 2008b, GRB Coordinates Network, 8159

\bibitem[Rol et al.(2008)]{2008GCN..8164....1R}
Rol, E., Tanvir, N., Rea, N., Wiersema, K., Skillen, I., \& Curran, P.~A.\ 2008, GRB Coordinates Network, 8164

\bibitem[Roming et al.(2005)]{2005SSRv..120...95R}
Roming, P.~W.~A., et  al.\ 2005, Space Science Reviews, 120, 95


\bibitem[\protect\citeauthoryear{Savchenko et al.}{2008}]{sav09}
Savchenko, V., et al. 2009, GRB Coordinates Network 8837

\bibitem[Snowden \& Schmitt(1990)]{1990Ap&SS.171..207S}
Snowden, S.~L., \& Schmitt, J.~H.~M.~M.\ 1990, ApJS, 171, 207


\bibitem[Str{\"u}der et al.(2001)]{2001A&A...365L..18S}
Str{\"u}der, L., et al.\ 2001, ApJ, 365, L18

\bibitem[Tanvir \& Varricatt(2008)]{2008GCN..8126....1T}
Tanvir, N.~R., \& Varricatt, W.\ 2008, GRB Coordinates Network, 8126


\bibitem[\protect\citeauthoryear{Thompson  \& Beloborodv}{2005}]{tb05}
Thompson C., \& Beloborodov, A.M., 2005, ApJ, 634, 565

\bibitem[\protect\citeauthoryear{Thompson  \& Duncan}{1993}]{td93}
Thompson C., \& Duncan, R.C., 1993, ApJ, 408, 194

\bibitem[Thompson \& Duncan(1995)]{1995MNRAS.275..255T}
Thompson, C., \& Duncan, R.~C.\ 1995, MNRAS, 275, 255


\bibitem[\protect\citeauthoryear{Thompson, Lyutikov \& Kulkarni}{2002}]{tlk02}
Thompson C., Lyutikov M., \& Kulkarni S.R., 2002, ApJ, 274, 332

\bibitem[Turner et al.(2001)]{2001A&A...365L..27T}
Turner, M.~J.~L., et al.\ 2001, ApJ, 365, L27

\bibitem[\protect\citeauthoryear{Ubertini et al.}{2003}]{ibis}
Ubertini, P., Lebrun, F., Di Cocco, G., et al. 2003, A\&A, 411, L131

\bibitem[Voges et al.(1992)]{1992eocm.rept..223V}
Voges, W., et al.\ 1992,  Environment Observation and Climate Modelling Through International Space
Projects, 223

\bibitem[\protect\citeauthoryear{von Kienlin \& Connaughton}{2009}]{vc09}
von Kienlin, A. \& Connaughton, V. 2009, GRB Coordinates Network 8838

\bibitem[\protect\citeauthoryear{Winkler et al.}{2003}]{integral}
Winkler, C., Courvoisier, T.J.-L., Di Cocco G., et al. 2003 A\&A, 411, L1

\bibitem[\protect\citeauthoryear{Woods et al.}{2004}]{woods04}
Woods P.M. et al. 2004, ApJ, 605, 378

\bibitem[\protect\citeauthoryear{Woods et al.}{2007}]{woods2}
Woods P.M. et al. 2007, ApJ, 654, 470


\bibitem[Woods et al.(2008)]{2008ATel.1824....1W}
Woods, P., Gogus, E.,  \& Kouveliotou, C.\ 2008, The Astronomer's Telegram, 1824



\end{thebibliography}
\end{document}